\documentclass[journal]{IEEEtran}

\usepackage{graphicx}
\usepackage{hyperref}

\usepackage{multirow, booktabs, siunitx,bm}
\usepackage{amsmath,amssymb,amsfonts}
\usepackage[nameinlink]{cleveref}
\usepackage{algorithmic}
\usepackage{graphicx}
\usepackage{textcomp}
\usepackage{booktabs} 
\usepackage{hyperref}
\usepackage{cleveref}
\usepackage{multirow}
\usepackage{color}
\usepackage{diagbox}
\usepackage{xcolor,colortbl}
\usepackage{enumitem}
\usepackage{balance}
\usepackage{soul}
\usepackage{color}
\usepackage{wallpaper}
\usepackage{graphicx}
\usepackage{cleveref}
\usepackage{numprint}
 \usepackage{rotating}
\usepackage{ amssymb }
\usepackage{float}
\usepackage{paralist}
\usepackage{multirow}
\usepackage[T1]{fontenc}
\usepackage{blindtext}

\newcommand{\eg}{e.\,g.,\,}
\newcommand{\ie}{i.\,e.,\,}

\newcommand{\cf}{{cf.\,}}

\usepackage{xspace}
\usepackage{soul}
\usepackage{color}

\usepackage[activate]{microtype}
\newcommand{\albert}{\textsc{AlBERT}\xspace}

\newcommand{\muse}{\textsc{MuSe-CaR\,}}

\begin{document}
\title{An Estimation of Online Video User Engagement from Features of Continuous Emotions} 

\author{Lukas Stappen,~\IEEEmembership{Member,~IEEE,}
        Alice Baird,~\IEEEmembership{Member,~IEEE,}
        Michelle Lienhart,
        Annalena B\"atz,
        and~\\Bj{\"o}rn Schuller,~\IEEEmembership{Fellow,~IEEE}
\IEEEcompsocitemizethanks{
\IEEEcompsocthanksitem All authors are with the EIHW -- Chair of Embedded Intelligence for Health Care and Wellbeing, University of Augsburg, Germany.\protect
\IEEEcompsocthanksitem Bj{\"o}rn Schuller is also with GLAM -- Group on Language, Audio, \& Music, Imperial College, London, UK.
\IEEEcompsocthanksitem Contact E-mail: {stappen,baird,schuller}@ieee.org}
}

\markboth{UNDER REVIEW}
{Shell \MakeLowercase{\textit{et al.}}: Bare Demo of IEEEtran.cls for IEEE Journals}



\maketitle


\begin{abstract}
Portraying emotion and trustworthiness is known to increase the appeal of video content. However, the causal relationship between these signals and online user engagement is not well understood. This limited understanding is partly due to a scarcity in emotionally annotated data and the varied modalities which express user engagement online. In this contribution, we utilise a large dataset of YouTube review videos which includes ca.\ 600 hours of dimensional arousal, valence and trustworthiness annotations. We investigate features extracted from these signals against various user engagement indicators including views, like/dislike ratio, as well as the sentiment of comments. In doing so, we identify the positive and negative influences which single features have, as well as interpretable patterns in each dimension which relate to user engagement. Our results demonstrate that smaller boundary ranges and fluctuations for arousal lead to an increase in user engagement. Furthermore, the extracted time-series features reveal significant ($p<0.05$) correlations for each dimension, such as, count below signal mean (arousal), number of peaks (valence), and absolute energy (trustworthiness). From this, an effective combination of features is outlined for approaches aiming to automatically predict several user engagement indicators. In a user engagement prediction paradigm we compare all features against semi-automatic (cross-task), and automatic (task-specific) feature selection methods. These selected feature sets appear to outperform the usage of all features, \eg using all features achieves $1.55$ likes per day (Lp/d) mean absolute error from valence; this improves through semi-automatic and automatic selection to $1.33$ and $1.23$ Lp/d, respectively (data mean $9.72$ Lp/d with a std. $28.75$ Lp/d).

\end{abstract}


%
\IEEEpeerreviewmaketitle

\section{INTRODUCTION}

\IEEEPARstart{O}{nline} video content hosted by platforms such as YouTube is now gaining more daily views than traditional television networks~\cite{Battaglio2016}. There are more than 2 billion registered users on YouTube, and a single visitor will remain on the site for at least 10 minutes~\cite{Cooper-19-YouTube}. Viewers rate of retention for a single video is between 70-80\%, and such retention times may be due to (cross-) social network effects \cite{yan2015youtube, 6521345, 8382331} and the overall improvement in content and connection quality in recent years \cite{lebreton2020predicting, dobrian2011understanding}, but arguably caused by intelligent mechanisms \cite{6522525}, \eg 70\% of videos watched on YouTube are recommended from the previous video~\cite{Cooper-19-YouTube}. To this end, gaining a better understanding of what aspects of a video a user engages with has numerous benefits~\cite{dobrian2011understanding}, including allowing a creator to predict the potential for a video to become \textit{viral} \cite{trzcinski2017predicting}, which may improve advertising opportunities and limit the spread of harmful misinformation~\cite{knuutila2020covid}. 

Positive emotion~\cite{berger2012makes} and trust of the individuals in videos ~\cite{nikolinakou2018viral} have shown to affect user (\ie content) engagement~\cite{shehu2016effects,kujur2018emotions}. In traditional forms of entertainment (\ie film) portraying emotion captivates the audiences improving their ability to remember details~\cite{subramanian2014emotion} and similar \textit{persuasion appeals} are applied within shorter-form YouTube videos~\cite{english2011youtube}. When emotion is recognised computationally, research has shown that the emotion (arousal and valence) of a video can be an indicator of popularity, particularly prominent when observing audio features~\cite{Sagha17-PTP}. 

The frequency of comments by users is also a strong indicator of how engaged or not users are with a video~\cite{yang2016video}. Furthermore, understanding the sentiment of comments (\ie positive, neutral, or negative) can offer further insights on the type of view engagement, \eg more positive sentiment correlates to longer audience retention~\cite{yang2016video}. 

In a similar way to the use of emotions, developing trust between the viewer (trustor) and the presenter (trustee) has also shown to improve user engagement. It is a common strategy by content creators to facilitate what is known as a \textit{parasocial relationship}. A parasocial relationship develops when the viewer begins to consider the presenter as a friend without having ever met them~\cite{chapple2017investigation}. 

With this in mind, we unite multiple emotional signals for an explicit engagement analysis and prediction in this current contribution. Thereby, we focus on the utilisation of the emotional dimensions of arousal and valence and extend the typical Russel circumplex model for emotion, by adding trustworthiness as a continuous signal. 
First, we aim to understand better continuous factors which improve metadata-related (\ie views, likes, etc.) and comment-related (\ie sentiment of comments, positive-negative ratios, likes of comments etc.) user engagement across modes (\ie emotional signals to text-based indicators). 
To do this, we collect the metadata as well as more than 75\,k comments from the videos. We annotate a portion of these comments to be used in combination with other data sets for training a YouTube comment sentiment predictor for the automatic assessment of the unlabeled comments. Furthermore, we utilise a richly annotated data set of ca.\ 600\,hours of continuous annotations \cite{stappen2021multimodal}, and derive cross-task features from this initial correlation analysis. Second, we compare these engineered, lean features, to a computationally intensive feature selection approach and to all features when predicting selected engagement indicators (\ie views, likes, number of comments, likes of the comments). We predict these indicators as a regression task, and train \textit{intrepretable} (linear kernel) Support Vector Regressors (SVR). 
To the best of the authors' knowledge, there has been no research which analyses YouTube video user engagement against trustability time-series features. Furthermore, we are the first to predict cross-modal user popularity indicators as a regression task -- purely based on emotional signal features without using typical text, audio, or images/video features as input. The newly designed and extended datasets, code, and the best models will be made publicly available on in our project repository\footnote{Github \& zenodo repositories to be made available upon acceptance in order to preserve originality.}.
\vspace{-0.2cm}
\section{BACKGROUND}

Within our contribution, the concept of emotions for user-generated content is extended from the conventional Russel concept of emotion dimensions, valence, and arousal~\cite{Russell}, to include a continuous measure for trustworthiness. In the following, we introduce these core concepts and related studies.

\vspace{-0.2cm}
\subsection{Concepts of emotion and trustworthiness} \label{sec:found_emotion}

There are two predominant views in the field of affective science: The first assumes that emotions are discrete constructs, each acting as an independent emotional system of the human brain, and hence, can be expressed by discrete categories~\cite{ekman1992argument}. The second assumes an underlying interconnected dimensional signal system represented by continuous affective states. 

For emotion recognition using continuous audio-video signals, the circumplex model of emotion developed by Russel is the most prominent~\cite{Russell} and applied~\cite{stappen2021multimodal, kossaifi2019sewa, busso2008iemocap} approach of the latter idea. This representation of affect typically consists of continuous valence (the positiveness/ negativity of the emotion) and arousal dimensions (the strength of the activation of the emotion), as well as an optional third focus dimension~\cite{posner2005circumplex}. 

In the past, both approaches to classify emotions in user-generated content~\cite{EmoClasYoTube} rely on Ekmann's model to predict six emotional classes in YouTube videos. Similarly,~\cite{zadeh2016mosi} annotated YouTube videos with labels for subjectivity and sentiment intensity~\cite{wollmer2013youtube} was the first to transfer the dimensional concept to YouTube videos. Recently,~\cite{kollias2019deep} annotated 300 videos (ca.\ 15 hours) of `in-the-wild' data, predominantly YouTube videos under the creative commons licence.

However, none of the mentioned datasets allows the bridging of annotated or predicted emotional signals with user engagement data from  videos. We fill this research gap utilising continuous emotional signals and corresponding data as well as providing insights into the novel dimension of trustworthiness, entirely without relying on word-based, audio, or video feature extraction. 

Although general literature lacks in providing a consistent concept of trustworthiness~\cite{horsburgh1961trust, moturu2011quantifying, cox2016trustworthiness}, in this work, we define trust as the ability, benevolence, and integrity of a trustee analogous to~\cite{colquitt2007trust}. In the context of user-generated reviews, the viewers assess from their perspective if and to what extent the reviewer communicates unbiased information. In other words, how truthful and knowledgeable does the viewer feel a review is at every moment? As we mentioned, building this trust is part of developing a parasocial relationship with the audience, and in doing so, likely increases repeated viewing~\cite{lim2020role}. 

\vspace{-0.2cm}
\subsection{Sentiment analysis of YouTube comments}
Sentiment Analysis studies the extraction of opinions, sentiments, and emotions (\eg ``positive'', ``negative'', or ``neutral'') of user-generated content. The analysed content usually consists of text~\cite{gilbert2014vader, boiy2007automatic}, such as in movie and product reviews as well as comments~\cite{siersdorfer2014analyzing, singh2013sentiment}. 
In recent years, the methods for text classification have developed rapidly. Earlier work using rule-based and classical word embedding approaches is now being replaced by what is known as \textit{transformer networks}, predicting \textit{context-based} word embeddings~\cite{devlin-etal-2019-bert}. State-of-the-art accuracy results on sentiment benchmark datasets using these methods~\cite{cuietal2019fine} range from 77.3 for the 3-classes twitter~\cite{nakovetal2013semeval} and between 72.4 and 75.0 on a 2-classes YouTube comment data sets~\cite{uryupinaetal2014sentube}.

In contrast to the literature, our approach utilises the predicted sentiment of a fine-tuned Word Embedding Transformer \albert~\cite{Lan2020ALBERT} to automatically classify comments on a large scale to investigate the cross-modal relationship to the continuous emotion and trustworthiness signals. 

\vspace{-0.2cm}
\subsection{Analysis of YouTube engagement data and cross-modal studies}
YouTube meta and engagement data are well researched \cite{yan2015youtube} with contributions exploring across domains \cite{6521345,8382331}, and focusing on both long \cite{6331531} and short form video sharing \cite{6522525,8305655}. 

Most previous work analyse view patterns, users' opinions (comments) and users' perceptions (likes/dislikes), and their mutual influence~\cite{bhuiyan2017retrieving}. Khan et. al. ~\cite{khan2014virality} correlated these reaction data, while~\cite{rangaswamy2016metadata} connects them to the popularity of a video.

An extended comment analysis has been conducted by~\cite{severyn2016multi} predicting the type and popularity towards the product and video. The comment ratings, thus the community acceptance, was predicted by~\cite{siersdorfer2010useful} using the comment language and discrete emotions. Moreover, in~\cite{wu2014correlation} the authors correlated  popularity measures and the sentiment of the comments. 
Data of other social networking platforms combine sentiment analysis and social media reactions~\cite{gilbert2014vader, ceron2014every},  and~\cite{preoctiuc2016modelling} attempted to map Facebook posts to the circumplex model to predict the sentiment of new messages.

To the best of our knowledge, no work has so far attempt to investigate the relationship to sophisticated continuous emotional and trustworthiness signals and based on these, predict user engagement as  regression tasks.

\vspace{-0.2cm}
\section{DATA}
\label{sec:data}

The base for our experimental work is the \muse data set\footnote{The raw videos and YouTube IDs are available for download: https://www.muse-challenge.org/challenge/get-data}\cite{stappen2021multimodal}. \muse is a multimedia dataset originally crafted to improve machine understanding of multimodal sentiment analysis in real-life media. For the first time, it was used for the \textsc{MuSe} 2020 Challenge, which aimed to improve emotion recognition systems, focusing on the prediction of arousal and valence emotion signals~\cite{stappen2020summary}. For a detailed description of typical audio-visual feature sets and baseline systems that are not directly related to this work, we point the reader to \cite{stappen2020muse}.

\vspace{-0.2cm}
\subsection{Video, meta- and engagement data}\label{sec:muse}
The dataset contains over 300 user-generated vehicle review videos, equal to almost 40 hours of material that cover a broad spectrum of topics within the domain. The videos were collected from YouTube\footnote{All owners of the data collected for use within the \muse data set were contacted in advance for the consent of use for research purposes.} and have an average duration of 8 minutes. The reviews are primarily produced by semi- (`influencers') or professional reviewers with an estimated age range of the mid-20s until the late-50s. The speech of the videos is English. We refer the reader to \cite{stappen2021multimodal} for further in-depth explanation about the collection, the annotator training, and the context of the experiments. Utilising the YouTube ID, we extend the data set by user engagement data. The explicit user engagement indicators are calculated on a per-day basis (p/d) as the videos were uploaded on different days resulting in views \textbf{(Vp/d)}, likes \textbf{(Lp/d)}, dislikes \textbf{(Dp/d)}, comments \textbf{(Cp/d)}, and likes of comments \textbf{(LCp/d)}. Per video the user engagement criteria is distributed ($\mu$\,mean, $\sigma$\,standard\,deviation) as; Vp/d: $\mu\,=\,863.88,\,\sigma\,=\,2048.43$; Lp/d: $\mu\,=\,9.73,\,\sigma\,=\,28.75$, Dp/d: $\mu\,=\,0.4125,\,\sigma\,=\,1.11$; Cp/d: $\mu\,=\,0.91, \sigma\,=\,3.00$; and LCp/d: $\mu\,=~5.28,\,\sigma\,=\,16.84$.

\vspace{-0.3cm}
\subsection{Emotion and trustworthiness signals}\label{sec:emotion}

%

As with emotions in general, a certain level of disagreement due to subjectivity can be expected~\cite{Russell}. For this reason, nine annotators were trained \cite{stappen2021multimodal} to have a common understanding of the arousal, valence, and trustworthiness concepts as discussed in~\Cref{sec:found_emotion}. As well established \cite{busso2008iemocap, kossaifi2019sewa}, the annotator moves the hand up and down using a \textit{Logitech Extreme 3D Pro Joystick} to annotate one of three dimensions, while watching the videos. The movements are recorded over the entire duration of the video sequence and sampled with a bin size of 0.25\,Hz on an axis magnitude between -1\,000 and 1\,000. Every annotation was checked by an auditor using quantitative and qualitative measures to ensure a high quality~\cite{baird2020considerations}. The time required for annotation alone stands for more than 600 working hours (40 hours video * 3 dimensions * 5 annotators per dimension).

\begin{figure}[t!]
    \centering
    \includegraphics[width=1.0\columnwidth]{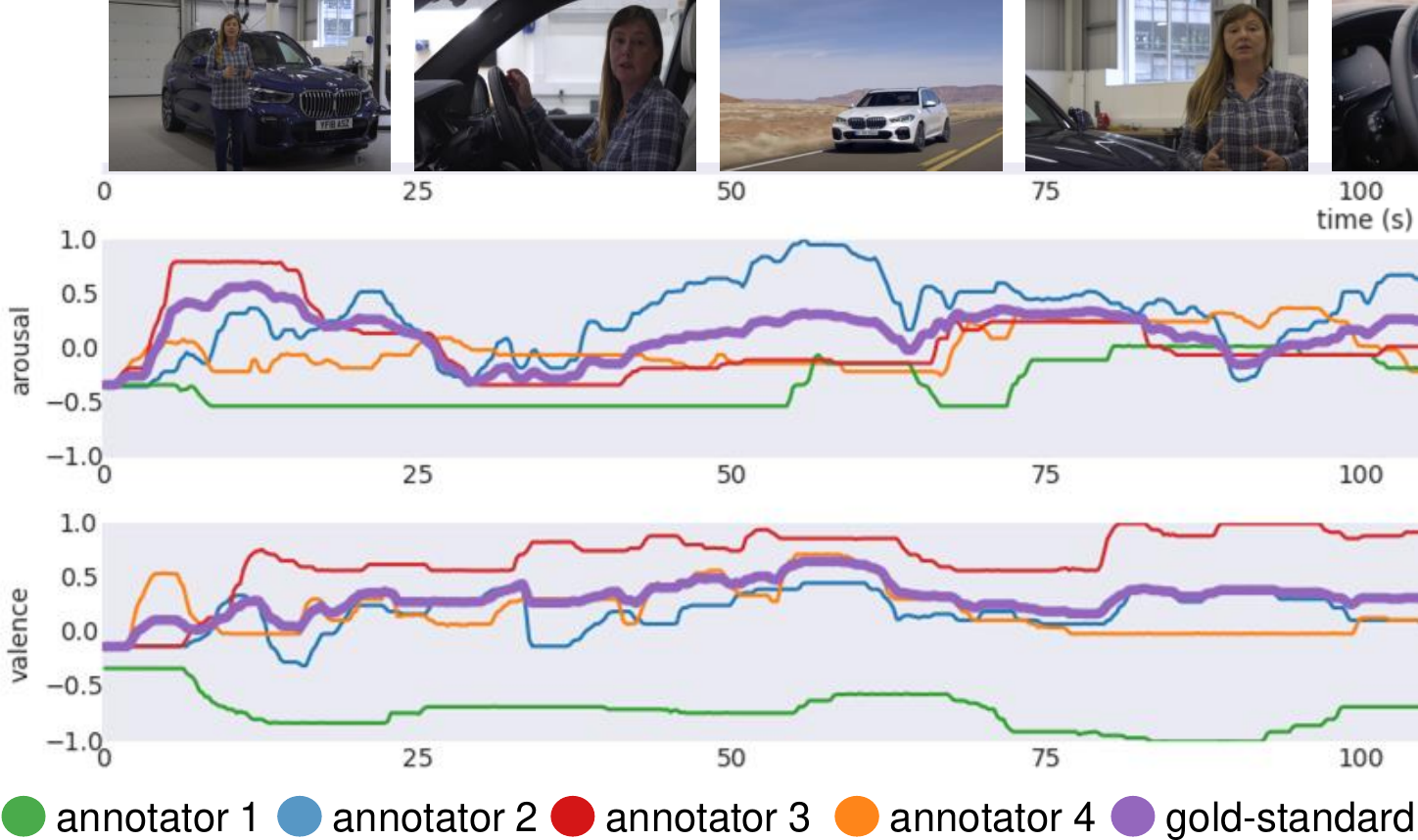} 
    \vspace{-0.4cm}
    \caption{Example video (video ID 5): Four annotation signals for arousal and valence in addition to the fused EWE ``gold-standard'' signal (bold purple). The annotator 1 (green) has a negative correlation to the others on valence. The weight ($r_1$) is set to 0, not considered for the fused gold-standard by EWE.} 
    \label{fig:dataset}
\end{figure}

The annotation of five independent annotators for each video and signal type are fused to obtain a more objective gold-standard signal as depicted in~\Cref{fig:dataset}.
For the fusion of the individual continuous signals, the widely established Evaluator Weighted Estimator (EWE) was computed~\cite{schuller2013intelligent,ringeval2017avec}. It is an estimator of inter-rater agreement, hence the personal reliability, in which the weighted mean corresponds to the calculated weights for each rater based on the cross-dependency of all other annotators. The EWE can be formulated as 
\begin{equation}
y_{n}^{E W E}=\frac{1}{\sum_{a=1}^{A} r_{a}} \sum_{a=1}^{A} r_{a} y_{n, a}, 
\end{equation}
where $y$ is a discrete point of the signal $n$ and $r_a$ is the reliability of the a-th rater. To use the data for later stages, we z-standardise them. 
\vspace{-0.3cm}
\subsection{Video comments}
\label{sec:video_comments}
Based on the video IDs of the corpus, we collected more than 79\,k YouTube comments and comment-related like counts excluding any other user information, such as the username. We focus exclusively on the parent comments, ignoring reaction from the child comments. The count of comment likes reflect the number of people sharing the same opinion and those who ``liked'' the comment. We randomly select 1\,100 comments for labelling, which is used as a quantitative estimator of how accurate our prediction of the other unlabelled comments are. Three annotators labelled each of them as positive, neutral, negative, and not applicable. The average inter-rater joint probability is 0.47. We use a majority fusion to create a single ground truth, excluding texts where no majority is reached.

\section{EXPERIMENTAL METHODOLOGY}
\label{sec:method}
\begin{figure*}[t!]
    \centering
    \includegraphics[width=0.7\linewidth]{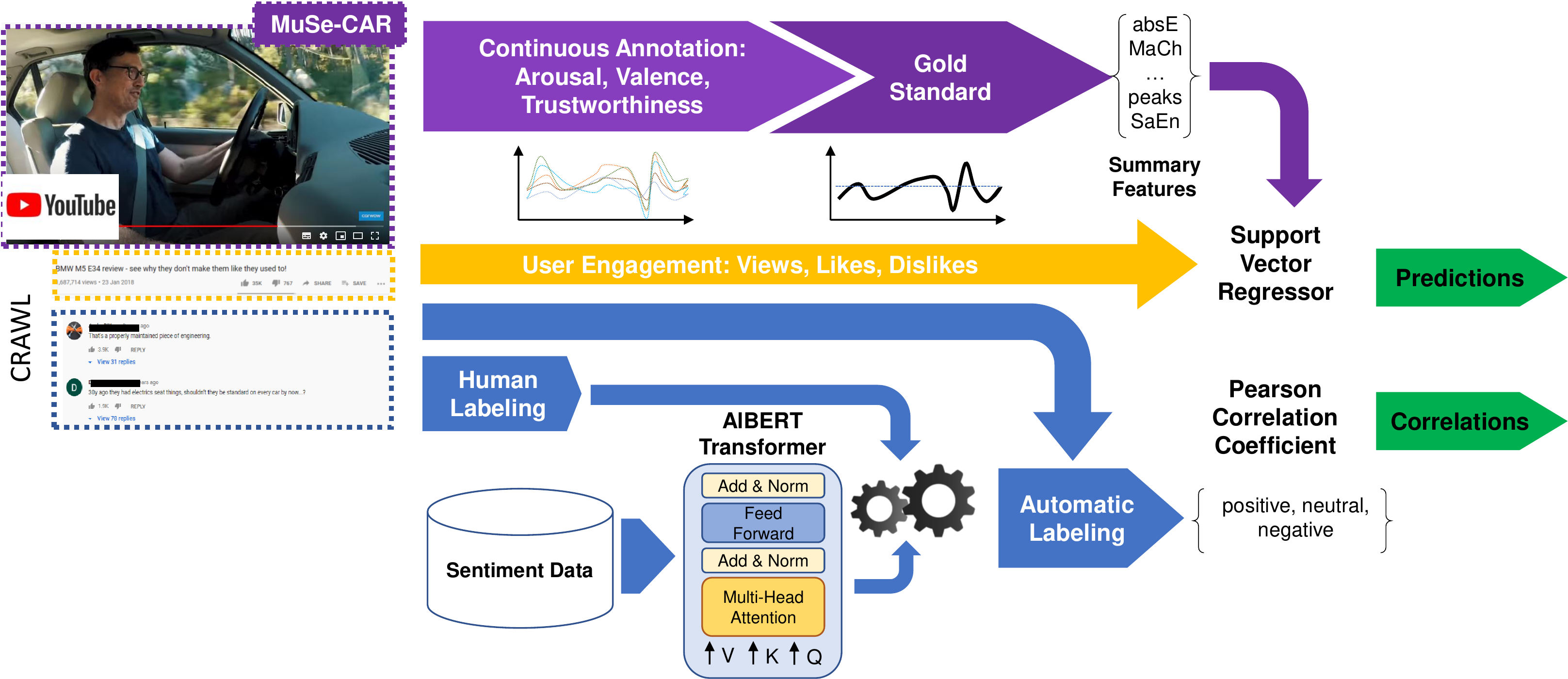}
    \caption{
    A comprehensive overview of the approach, as explained in \Cref{sec:method}. Statistics derived from continuous emotion and trustworthiness (purple), automatic sentiment labelling of YouTube comments (blue), and user engagement data (orange) are preprocessed to investigate correlations and predict engagement from features. The development of our signal feature extraction as a cornerstone of our analysis is described in detail in \Cref{sec:featureext}. With its help, we aim to uncover relationships to engagement data  (\cf \Cref{sec:corr_measure}) and the sentiment of the comments predicted by our trained network (\cf \Cref{sec:sentcomments}). They also serve as input to our regression experiments to predict user engagement (\cf \Cref{sec:training}).}
    \label{fig:overview}
\end{figure*}

\Cref{fig:overview} gives an overview of our approach. As a cornerstone of our analysis (\cf \Cref{sec:featureext}), annotation of arousal, valence, and trustworthiness are annotated by five independent annotators. These signals are then fused (\cf \Cref{sec:emotion}) to a gold standard label, and meaningful features are extracted (\textbf{purple}). In addition, YouTube user engagement-related data (\textbf{yellow}) and the comments are scraped (\textbf{blue}) from each video. Several sentiment data sets are collected and merged in order to train a robust sentiment classifier using a Transformer network \albert to predict unlabelled YouTube comments after fine-tuning on several datasets and our labelled comments.
Then, we first investigate correlations between the predicted sentiment of the YouTube comments, the YouTube metadata, and the statistics derived from the continuous signals (arousal, valence, and trustworthiness). Additionally, we use derived features to predict user engagement (Vp/d, Lp/d, Cp/d, CLp/d) directly.

\vspace{-0.3cm}
\subsection{Feature extraction from signals}
\label{sec:featureext}

A signal is usually sampled to fine-grained, discrete points of regular intervals, which can be interpreted as a sequential set of successive data points over time~\cite{adhikari2013introductory}. Audio, video, and psychological signals are widely used for computational analysis~\cite{schuller2013intelligent, schuller2020interspeech}. Simple statistics and advanced feature extraction can be applied in order to condense these signals to meaningful summary representations and make them more workable~\cite{christ2018time}. In this work, we use common statistical measures such as the standard deviation (\textit{std}), and 5\%-, 25\%-, 50\%-, 75\%-, and 95\%-\textit{quantiles} (\begin{math}q_{5}, q_{25}, q_{50}, q_{75}, q_{95}\end{math}) as they are less complex to interpret, and have been applied in related works \cite{Sagha17-PTP}. Furthermore, to make better use of the changes over time, we manually select and calculate a wide range of time-series statistics following previous work in similar fields~\cite{geurts2001pattern, schuller2002automatic}. For example, in computational audition (\eg speech emotion recognition), energy-related features 
of the audio signals are used to predict emotions~\cite{schuller2002automatic}.

We calculate the dynamic sample skewness (\textit{skew}) of a signal using the adjusted Fisher-Pearson standardised moment coefficient, to have a descriptor for the asymmetry of the series
~\cite{doane2011measuring,ekman1992argument}.
Similarly, the kurtosis (\textit{kurt}) measures the `flatness' of the distribution by utilising the fourth moment~\cite{westfall2014kurtosis}. 
Of the energy-related ones, the absolute energy (\textit{absE}) of a signal can be determined by the sum over the squared values~\cite{christ2018time}.
\vspace{-0.1cm}
\begin{equation}
absE=\sum_{i=1, \ldots, n} x_{i}^{2},
\end{equation}
where $x$ is the signal at point $i$.
Also well known for physiological time-series signals is the sample entropy (\textit{SaEn}), a variation of the approximate entropy, to measure the complexity independently of the series length~\cite{yentes2013appropriate, richman2000physiological}. Several change-related features might be valuable to reflect the compressed signal~\cite{christ2018time}: 
First, the sum over the absolute value of consecutive changes expresses the absolute sum of changes (\textit{ASOC}):
\begin{equation}
ASOC = \sum_{i=1,...,n-1}|x_{i+1}-x_{i}|.
\end{equation}
Second, the mean absolute change (\textit{MACh})  over the absolute difference between subsequent data points is defined as: 
\begin{equation}
MACh = \frac{1}{n}  \sum_{i=1,...,n-1}|x_{i+1}-x_{i}|,
\end{equation}
where $n$ is the number of time-series points. Third, the general difference between consecutive points over time is called the mean change (\textit{MCh}):
\begin{equation}
MCh = \frac{1}{n-1}  \sum_{i=1,...,n-1}x_{i+1}-x_{i}.
\end{equation}
Fourth, the mean value of a central approximation of the second derivatives (\textit{MSDC}) is defined as:
\begin{equation}
MSDC = \frac{1}{2*(n-1)}  \sum_{i=1,...,n-1}0.5\cdot(x_{i+2}-2\cdot{i+1}+x_{i}).
\end{equation}
Finally, the length of the normalised consecutive sub-sequence is named strike above (\textit{LSAMe}) and below (\textit{LSBMe}) the mean. To summarise the distribution similarity, the normalised percentage of reoccurring datapoints (\textit{PreDa}) of non-unique single points can be calculated by taking the number of data points occurring more than once divided by the number of total points. Also early or late high and low points of the signal are of descriptive value. Four single points describe these: the first and last location of the minimum and maximum (\textit{FLMi}, \textit{LLMi}, \textit{FLMa}, and \textit{FLMa}) relatively to the length of the series. The last two count a) the number of crossings of a point \textit{m} (here: \textit{m=0}) (\textit{CrM}), where for two successive time series steps are first lower (or higher) than \textit{m} followed by two higher (or lower) ones~\cite{christ2018time} and b) the \textit{peaks} of the least support \textit{n}. A peak of support \textit{n} is described as a subsequence of a series where a value occurs, bigger than its \textit{n} neighbours to the left and the right~\cite{christ2018time,palshikar2009simple}. In total, we extract 24 features from one signal.
\vspace{-0.2cm}
\subsection{Sentiment extraction from comments}\label{sec:sentcomments}
\label{model_settings}

Given the vast amount of comments, we decided to carry out the labelling of the sentiment automatically and label only a small share of them by hand to quantify the prediction quality (\cf \Cref{sec:video_comments}). For this reason, we built a robust classifier for automatic YouTube sentiment prediction using PyTorch. We opted to use \albert as our competitive Transformer architecture~\cite{Lan2020ALBERT}. Compared to other architectures, \albert introduces two novel parameter reduction methods: First, the embedding matrix is separated into two more compact matrices, and second, layers are grouped and used repeatedly. Furthermore, it applies a new self-supervised loss function that improves training for downstream fine-tuning tasks. These changes have several advantages, such as reducing the memory footprint, accelerating the converge of the network, and leading to state-of-the-art results in several benchmarks~\cite{devlin-etal-2019-bert}. 

Before training, we remove all words starting with a ``\#``, ``@`` or ``http`` from all text sources and replace emoticons' unicode by the name. We train \albert in a two-step procedure. First, we fine-tune the model for the down-stream task of general sentiment analysis. No extensive YouTube comment data set is available, which would span the wide range of writing styles and expressed opinions. Therefore, we aggregate several datasets which aim to classify whether a text is positive, negative, or neutral as our initial training data: all data sets from SemEval (the Semantic Evaluation challenge), a series of challenges for computer-based text classification systems with changing domains ~\cite{nakovetal2013semeval} \eg Twitter, SMS, sarcasm, from 2013 to 2017 consisting of more than 76\,k data points; the popular US Airline Sentiment data set ~\cite{AirlineDataset} (14.5\,k tweets), and finally, 35\,k positive and 35\,k negative text snippets are selected from Sentiment140~\cite{go2009twitter}. The 60\,k positive, 32\,k neutral, and 56\,k negative text snippets are equally stratified and partitioned into 80-10-10 splits for training. We provide this selection for reproducibility in our code. 

Following the authors' recommendation, \albert is trained using a learning rate of \textit{1e-5}, a warmup ratio of \textit{0.06}, $\epsilon$ set to \textit{1e-8}, and gradient clipping set at \textit{1.0}. In addition, we use half-precision training and a batch size of 12 to fit the GPU memory restrictions (32\,GBs). Counteracting adverse effects of class imbalance, we further inject the class weight to each data point. The model converges after three epochs.
Next, we use our own YouTube comment data set to validate the results and further fine-tune the model. This version is then further trained in a second fine-tuning step using the 60\% of the YouTube comments and a reduced learning rate of \textit{1e-6} for one epoch.

The relative distribution of the classified sentiment of the YouTube comments is given in~\Cref{tab:sentiment_dist}. The model achieves an f1 score on the development of 81.13\% and 78.09\% on the test partitions as well as 75.41\% on the sample of our crawled and manually labelled YouTube test set.

\begin{table}[t!]
\centering
\caption{Example comments and sentiment distribution within the YouTube comments predicted by our developed sentiment model.}
\resizebox{\linewidth}{!}{%
    \begin{tabular}{l|ccp{6.2cm}}
    \toprule
    \textbf{sentiment} & \textbf{\# comments} & \textbf{predicted [\%]} & \textbf{example}\\
    \hline
    \hline
    positive & 26\,032  & 33 & ``the metaphors are just flying  like the raindrops in this video.`` \#47620 \\
    neutral &  28\,518 & 36 & ``Are engines for F30 made in Germany?''  \#4 \\
    negative & 24\,494 & 31 & ``Poor review unfortunately, the microphone  quality was very muffled...`` \#31 \\
    \bottomrule
    \end{tabular}
\label{tab:sentiment_dist}}
\end{table}
\vspace{-0.2cm}



\vspace{-0.2cm}
\subsection{Correlation measure and significance}\label{sec:corr_measure}

The Pearson correlation ($r$) explores the relationship between two continuous variables~\cite{ahlgren2003requirements}. Thereby, the relationship has to be linear, meaning that when one variable changes, the other also changes proportionally. $r$ is defined by

\begin{equation}
r_{x,z} = \frac{cov(x,z)}{\sigma_{x}\cdot\sigma_{z}} = \frac{\sum_{i=1}^{n} (x_{i}-\bar{x})\cdot(z_{i}-\bar{z})} 
{\sqrt{\sum_{i=1}^{n}(x_{i}-\bar{x})^2\cdot\sum_{i=1}^{n}(z_{i}-\bar{z})^2}}, 
\end{equation}

where $\mathrm{cov}(x, z)$ is the co-variance, a measure of the joint variability, of the variables $X$, $Z$, and $\sigma_{x}$, $\sigma_{z}$ -- the standard deviations of both variables~\cite{surhone2010spearman}. The resulting correlation coefficient lies between $-1$ and $+1$. If the value is positive, the two variables are positively correlated. A value of $(+/-)1$ signifies a perfect positive or negative correlation. A coefficient equal to zero implies that there is no linear dependency between the variables. 

For significance testing, we first compute the t-statistic, and then twice the survival function for the resulting t-value to receive a two-tailed p-value, in which the null hypothesis (two variables are uncorrelated) is rejected at levels of \begin{math}\alpha=\{0.01, 0.05,0 .1\}\end{math}~\cite{sham2014statistical}. 

\subsection{Feature selection}
To the best of our knowledge, we are the first extracting advanced features directly from emotional signals. Usually, not all engineered features are equally relevant. Since no previous research can guide us to a reliable selection, we propose two ways for feature selection for our task of predicting user engagement. The first is a correlation-based, \textit{cross-task semi-automatic selection} that uses the correlation between the feature and the target variables. Only those features whose mean value over all prediction tasks is between $ -0.2 > r_{mean} > +0.2$ (minimum low positive/negative correlation) are selected.

The other concept is a regression-based, \textit{task-specific automatic selection} with three steps. First, univariate linear regression ($f$) tests act as a scoring function and run successively to measure the individual effect of many regressors:

\begin{equation}
score(f,y) = \frac{X_{k_{i}} - \bar{X_{k_{i}}} \cdot (y - \bar{y}) } {\sigma_{X_{k_{i}}} \cdot \sigma_y}, 
\end{equation}
where $k_i$ is the feature index. The score is converted to an F-test estimate and then to a p-value. Second, the highest $k$ number of features are selected based on the p-value. Finally, this procedure runs brute-force for all number of feature combinations, where $5 < k < k_{max-1}$.

\subsection{SVR training procedure} \label{sec:training}
For our regression experiments, we use a Support Vector Regression (SVR) with a linear kernel as implemented by the open-source machine learning toolkit Scikit-Learn~\cite{scikit-learn}. The linear kernel allows us to interpret the weights from our various feature selections and has, among other applications, found wide acceptance in the selection of relevant genes from micro-array data~\cite{guyon2002gene}. Since the coefficients are orthogonal to the hyper-plane, a feature is useful to separate the data when the hyper-plane is orthogonal to the feature axis. The absolute size of the coefficient concerning the other features indicates the importance.

The training is executed on the 60-20-20 training/development/test partition split partitions, pre-defined in the \muse emotion recognition sub-task~\cite{stappen2020summary} (\cf \Cref{sec:muse}). During the training phase, we train a series of models on the training set with different parameters $C$ $\in$ \{$10^{-7}$, $10^{-6}$, $10^{-5}$, $10^{-4}$, $10^{-3}$, $10^{-2}$, $10^{-1}$, $1$ \} up to 10\,000 iterations and validate the performance on the development set. The best performing $C$ value is then used to re-train the model on an enlarged, concatenated training and development set, to estimate the generalisation performance on the hold-out test set. This method is repeated for each input signal (combination) on each target (\%). Due to the various scales of the input features, we apply standardisation to the data but leave the targets, as they allow interpretability of the results. The prediction results are evaluated using the Mean Absolute Error (MAE). 

\section{RESULTS AND DISCUSSION}\label{sec:results}

\begin{figure}[t!]
    \centering
    \includegraphics[width=0.8\columnwidth]{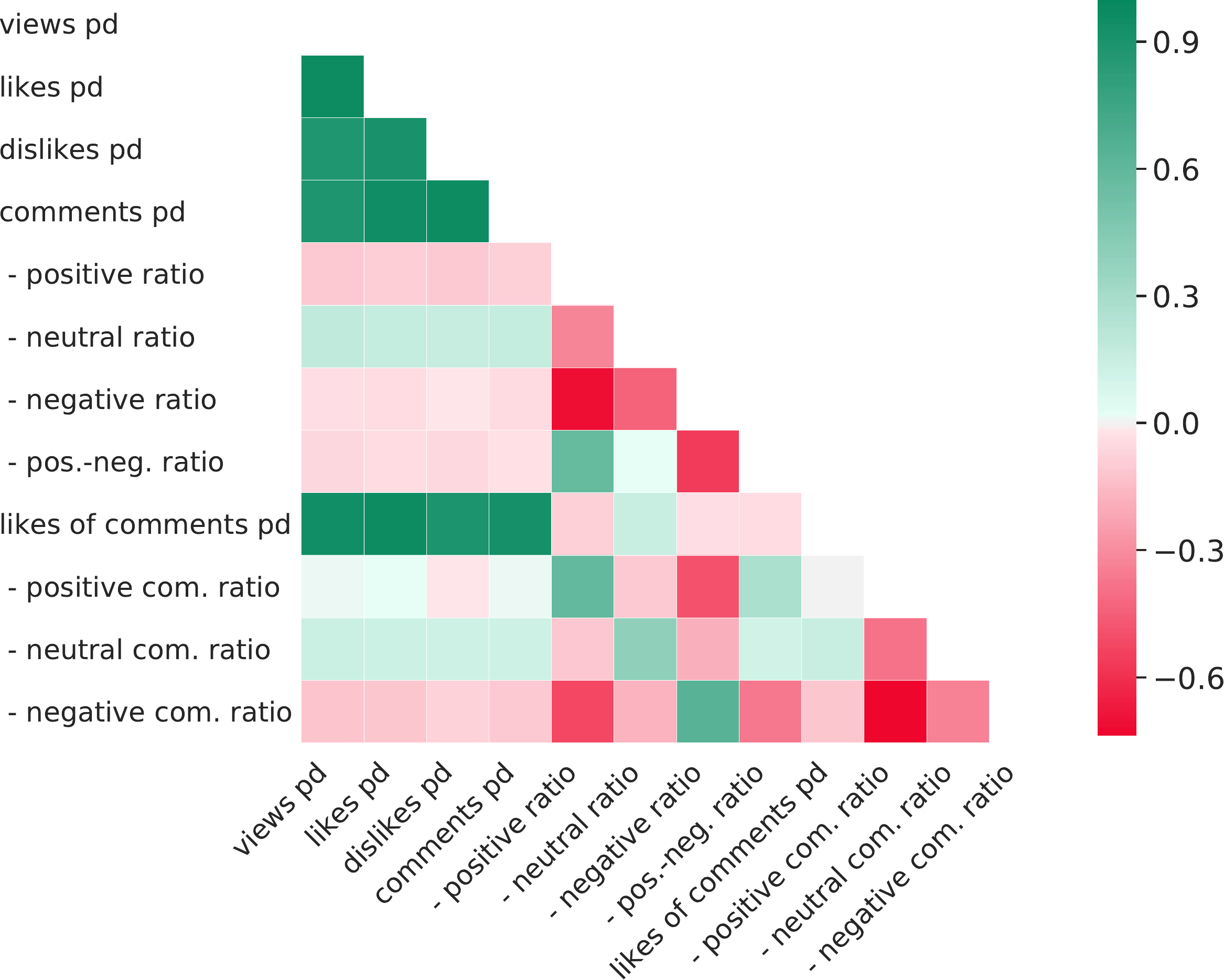} 
    \vspace{-0.2cm}
    \caption{Pearson correlation matrix of metadata. All results are considered to be significant at a 0.01 level. Abbreviations: comment (com.); positive (pos.); negative (neg.); per day (pd).
    }
    \label{fig:overall}
\end{figure}

\Cref{fig:overall} depicts the Pearson correlations for the user engagement indicators, and we see that the number of Vp/d, Lp/d, Dp/d, and Cp/d are highly correlated. Furthermore, when a correlation to one of these variables occurs, it is likely to be accompanied by correlations to others. 
We would like to note that the average relationship between likes and dislikes in our crawled videos is not as antagonistic as one might expect, which means that as the number of Lp/d increases; so to does the Dp/d, this may relate to the topic of the dataset, being that it is review videos, and the like or dislike may be more objective than other video themes.


\vspace{-0.3cm}
\subsection{Relationship between features and user engagement}

\begin{figure*}
    \centering
    \includegraphics[width=1\linewidth]{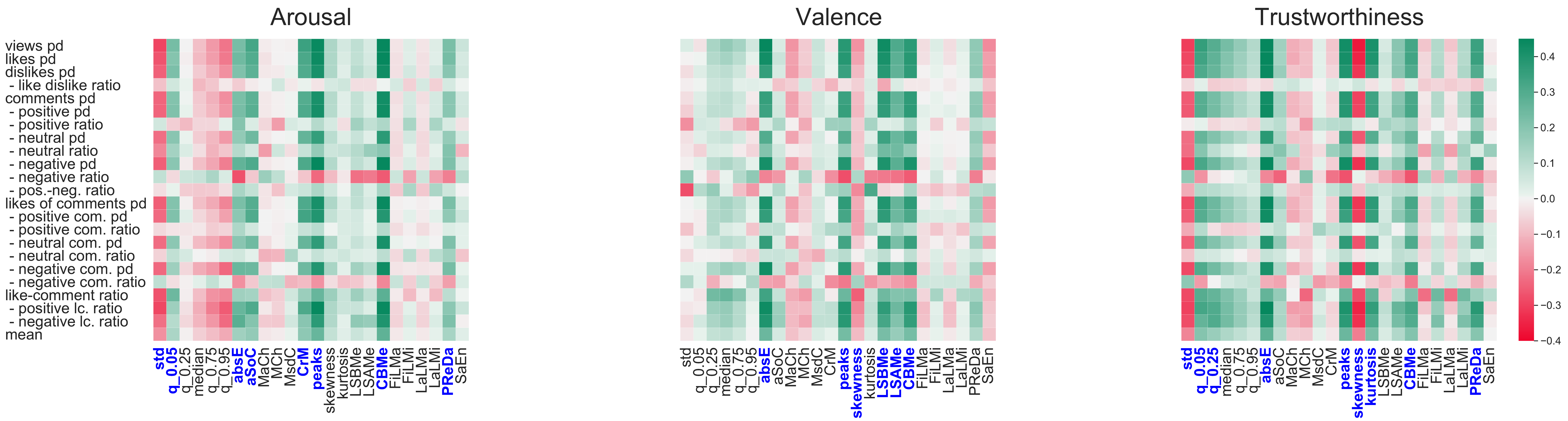}
    \label{fig:correlation}
    \vspace{-0.5cm}
    \caption{Pearson correlation matrix of user engagement indicators and the statistics/ features extracted from each dimension. The latter are standard deviation ($std$), quantile ($q_x$), absolute energy (\textit{absE}), mean absolute change (\textit{MACh}), mean change (\textit{MCh}), mean central approximation of the second derivatives (\textit{MSDC}), crossings of a point $m$ (\textit{CrM}), peaks, skewness, kurtosis, strike above the mean (\textit{LSAMe}), strike below the mean (\textit{LSBMe}), count below mean (\textit{CBMe}), absolute sum of changes (\textit{ASOC}), first and last location of the minimum and maximum (\textit{FLMi}, \textit{LLMi}, \textit{FLMa}, \textit{FLMa}), perc.\ of reoccurring datapoints ($PreDa$), and sample entropy (\textit{SaEn}). Features in blue are utilised as cross-task, semi-automatic features for user engagement prediction.}
    \vspace{-0.2cm}
\end{figure*}

Within this section, we discuss the correlation results for each emotional dimension seperately. We report Pearson correlation coefficients, as depicted in \Cref{fig:correlation}. Detailed results ($r$ and significance level) can be found in \Cref{fig:correlation}.

\textbf{Arousal: }The statistics extracted from the arousal signal indicate several correlations to the engagement data. When the $standard\,deviation$ or the level of the $quantile_{.95}$ increases, the number of Vp/d, Lp/d, Cp/d, and CLp/d slightly decreases (\eg \begin{math}r_{(views,std)}=-.293, r_{(views,q_{95})}=-.212\end{math}) with direct effect on the comment-like ratio (clr), \eg \begin{math}r_{std}=-.271\end{math}. In contrast, the level of the $quantile_{.05}$ has the opposite effect on all these metrics (\eg \begin{math}r_{(views,q_{.05})}=0.231, r_{(clr,q_{.05})}=-.248\end{math}. 
Of the more complex time-series statistics, the $peaks$ as well as the $CBM$ have the strongest correlations across most indicators. These indicates a moderate positive linear relationship, for instance, to Vp/d and Lp/d \begin{math}r_{(views, peaks)}=.440, r_{(likes,CBMe)}=.456\end{math} as well as Cp/d \begin{math}r_{peaks}=.409\end{math}. Further, when these features increase, the share of neutral comments increases much less than the share of positive and negative comments. 
The next strongest correlated features, $CrM$, $aSoc$, $abE$, and $PreDa$, also represent upward correlation slopes to the user-engagement criteria. Although these features reflect the general change in engagement, no conclusions can be drawn regarding sentiment of the engagement, as there is no significant correlation of any feature to the ratios (\eg like-dislike, and positive-negative comments).

\textbf{Valence: }Most statistics of the signal distribution are below $r = .2$, suggesting that there are only very weak linear dependencies with the engagement indicators. 
The only exceptions is the positive-negative ratio for the comments (\begin{math}r=-.276\end{math}) -- a lower $standard\,deviation$ leads to an increase in the proportion of positive comments. Furthermore, higher values around the centre of the distribution (kurtosis -- \begin{math}r=-.313\end{math}) to more likes per comment.
The strongest positively correlated feature is $absE$ \eg \begin{math}r_{views}=.467, r_{likes}=.422, r_{dislikes}=.355, r_{comments}=.350\end{math}, followed by the $peaks$, $CBME$ and $LSBMe$, which suggest the greater the value of these features, the greater the user engagement.
In contrast, the \textit{MaCh} and the \textit{SaEn} have significant slight negative correlations, which implies that when the valence signal of a video has a high complexity, the video has a higher tendency to receive fewer user engagement. 

\textbf{Trustworthiness: }The higher the level of $quantile_{.05}$, $quantile_{.25}$, $median$, and $quantile_{.75}$ (all slightly positively correlated, with decreasing relevance \eg \begin{math}r_{(views,q_{.05})}=.356, r_{(likes,q_{.75})}=.175\end{math}), the higher the Vp/d, Lp/d, Dp/d, Cp/d, and CLp/d. 
Similar to the valence dimension, we see that there is a negative effect on these engagement indicators when the standard deviation in the trustworthiness signal is higher \eg \begin{math}r_{(views,std)}=-.304, r_{(likes,std)}=-.287\end{math}.  
As for the other features, the $absE$ and the number of $peaks$ have a moderate positive correlation. The $skewness$ shows a significant negative correlation above $r < -.3$ for most indicators. In other words, a negative $skew$ of the trustworthiness signal, when the mass of the distribution is concentrated to the right (left-skewed), has a positive influence on user engagement.
Regarding the positiveness/negativeness sentiment ratios (like-dislike, comments positive-negative ratio), none of the features show significant associations.

\textbf{Result Discussion: }When observing the results from the above sections, we see several patterns between the emotion (including trust) signal statistics and user engagement. While the standard statistics of arousal show that bounded arousal (higher lower quantiles and lower high quantiles) and higher trustworthiness scores (all quantiles are positively correlated, with lower quantiles at a higher level) leads to more user engagement, the sentiment of a video seems less influential contrary to the findings of \cite{Sagha17-PTP}. Regarding the time-series features, the number of peaks with support \textit{n} = 10 seems a stable indicator across all signals. The energy-related features of valence and trustworthiness (\begin{math}valence = r_{(views,absE)}=.467, trustworthiness = r_{(views,absE)}=.497\end{math}) seem to have a medium-strong relationship and most likely a valuable predictive feature. 

Regarding the comments, independently of the type of signal and statistic, the negative comments seem to be higher correlated consistently, followed by the number of likes and positive comments. Overall, mostly slight to modest correlations are found. However, significant correlations, especially to the more complex time-series features, between valence, arousal, and trustworthiness levels in a video to the user engagement (number of users who watch it, like it, dislike it, or leave a comment) is evident.


\subsection{Predicting user engagement from features of emotion and trustworthiness signals}

~\Cref{tab:regression} shows the results of the prediction tasks Vp/d, Lp/d, Cp/d,  and CLp/D. It is worth noting that the scores vary according to the underlying scale of the target variables (\cf ~\Cref{sec:data}). 

\begin{figure}[t!]
    \centering
    \includegraphics[width=1\columnwidth]{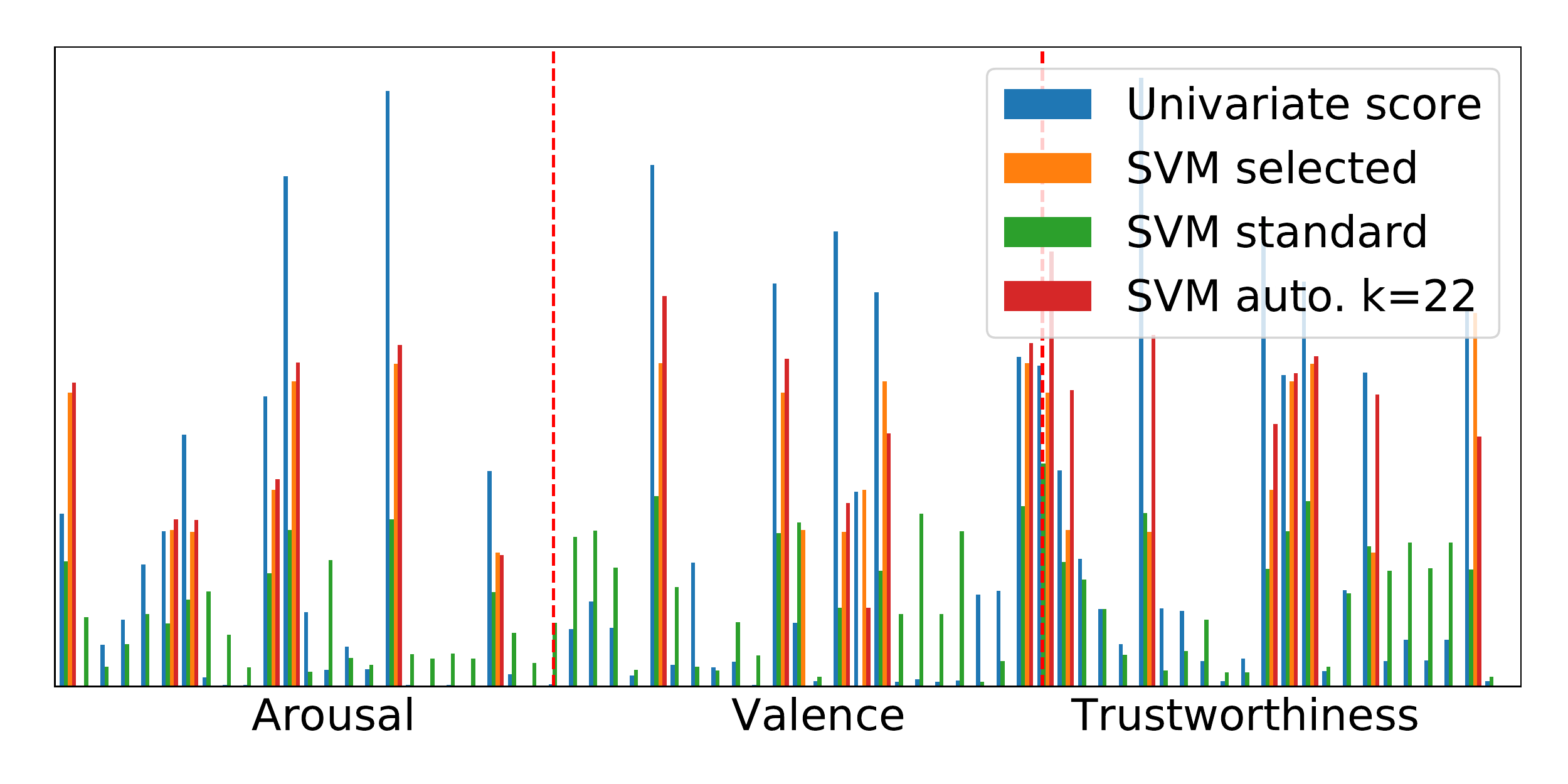} 
    \vspace{-0.9cm}
    \caption{SVM weights of arousal, valence, and trustworthiness features predicting the likes of comments comparing the use of all features, manually selected (24), and automatically selected $k$ = 22 features. To receive a real output and fit the p-values of the automatic selection in scale, we apply a base 10 logarithm and divide the result by 10 ($-Log(p_{value})/10$).} 
    \label{fig:svmweights}
    \vspace{-0.6cm}
    
\end{figure}

The features utilised from the cross-task semi-automatic feature selection method are highlighted (in blue) in \Cref{fig:correlation} for each feature type. Across the seven experiments the automatic selection process selected on average the following number of features per each criteria; $7.6$ Vp/d, $23.3$ Cp/d, $29.3$ Lp/d, and $20.1$ LCp/d. For each dimension, an average of $9.3$ for arousal, $9.5$ for valence, and $6.0$ for trustworthiness was selected. 
\Cref{fig:svmweights} illustrates an example of both selection methods for predicting CLp/d from a fusion of all three feature types. The p-values of the automatic (univariate) selection and the corresponding weights of all resulting SVMs are shown, indicating the relevance of each feature for the prediction.
The most informative features (largest p-values) also receive most weight from the corresponding SVM, indicating that the automatic selection is sensible. In this particular case, the hand selected features have almost identical weights as the automatic ones, whereby the missing features are enough to make the results worse than in the case of the other two (\cf \Cref{tab:regression}), indicating a high sensitivity if certain features are left out. 

\textbf{Views Per day: }When observing the Vp/d prediction from all features, we obtain the best result when performing an early fusion of the valence and trustworthiness signals, and with the addition of arousal, there is a minor decrease (205.8 and 205.8 MAE respectively); this demonstrates the predictive potential of all signals.
However, when applying our semi-automatic cross-task feature selection, there is a more substantial improvement particularly for arousal and valence as mono signals, obtaining $198.5$, and $184.8$ MAE, respectively. This improvement is increased further for valence through automatic feature selection, with our best results for Vp/d of $169.5$ MAE. Feature selection appears in all cases to not be beneficial for fused features, with arousal and valence improving slightly but no more than if the signal was alone. Without any feature selection trustworthiness is our strongest signal, for further investigation exploreing why trustworthiness does not improve at all with either of the feature selection methods ($218.7$ and $228.0$, for sel. and auto., respectively) would be of interest. 

\begin{table*}[t!]

\caption{
Prediction of views, likes, comments, and likes of comments aggregated per day utilising features extracted and crafted from \textbf{Arousal} (A), \textbf{Valence} (V), and \textbf{Trustworthiness} (T). 
We report $C$: parameter of the SVR, optimised for from $0.00001$ to $1$, using the best $M$: mean absolute error on the dev(elopment) set to define $C$ for test set prediction. (\%) indicates the relative change of the automatic (auto.) and semi-automatically selected (sel.) in \% to the unchanged features, ``+'' indicates an improvement, thus a decrease of the MAE compared to the original feature sets.}
\resizebox{1.0\linewidth}{!}{ 

\begin{tabular}{c||rrr|rrr||rrr|rrr||rrr|rrr||rrr|rrr}
\toprule
  \multirow{4}{*}{\textbf{Type}} & \multicolumn{6}{c||}{\textbf{Views}} & \multicolumn{6}{c||}{\textbf{Likes}} & \multicolumn{6}{c||}{\textbf{Comments}} & \multicolumn{6}{c}{\textbf{Likes of Comments}} \\
             & \multicolumn{3}{c}{\textbf{dev}} &        \multicolumn{3}{c||}{\textbf{test}}  &        \multicolumn{3}{c}{\textbf{dev}} &         \multicolumn{3}{c||}{\textbf{test}} &         \multicolumn{3}{c}{\textbf{dev}}  &       \multicolumn{3}{c||}{\textbf{test} } &              \multicolumn{3}{c}{\textbf{dev}} & \multicolumn{3}{c}{\textbf{test}} \\
        
&all&sel.&auto.&all&sel.&auto.&all&sel.&auto.&all&sel.&auto.&all&sel.&auto.&all&sel.&auto.&all&sel.&auto.&all&sel.&auto. \\
&MAE&rel.\%&rel.\%&MAE&rel.\%&rel.\%&MAE&rel.\%&rel.\%&MAE&rel.\%&rel.\%&MAE&rel.\%&rel.\%&MAE&rel.\%&rel.\%&MAE&rel.\%&rel.\%&MAE&rel.\%&rel.\% \\   
\hline
\textbf{A}& 231.8& +6.8& +5.0& 220.3& +9.9& +3.1& 2.30& -0.3& +2.6& 1.55& +5.9& +3.0& .288& -0.1& +3.7& .154& +2.5& +0.6& 1.19& +5.7& +5.9& .50& -19.1& -22.7 \\
\textbf{V}& 253.1& +8.7& +7.2& 223.8& +17.4& +24.3& 2.29& +0.6& +1.0& 1.61& +17.6& +24.0& .288& +3.1& +3.9& .154& +5.1& +2.4& 1.17& -1.4& +3.6& .51& -2.8& -18.4  \\
\textbf{T}& 237.4& +11.8& +16.3& 207.9& -5.2& -9.7& 2.21& +5.3& +14.4& 1.92& +13.3& +3.6& .262& +5.8& +6.4& .225& +2.1& -5.3& 1.11& -0.1& +9.5& .75& +8.8& +6.7 \\
\textbf{A+V}& 237.6& -1.0& +2.1& 210.7& +4.1& +18.3& 2.27& -11.4& +0.3& 1.79& +24.2& -0.7& .277& -4.3& +3.4& .161& +9.9& +0.1& 1.16& +0.1& +2.0& .54& +16.8& -27.3 \\
\textbf{A+T}& 240.3& +9.2& +15.7& 207.9& -6.7& -3.9& 2.26& +4.8& +10.6& 2.02& +11.8& +10.3& .268& +1.6& +7.2& .182& -34.9& -1.1& 1.11& -0.2& +3.7& .59& -17.9& -14.7 \\
\textbf{V+T}& 249.1& +15.5& +20.0& 205.8& -3.1& -2.6& 2.07& -11.8& -0.2& 1.99& +17.2& +0.1& .262& -2.7& +5.5& .188& +10.9& -24.7& 1.04& -6.2& +0.3& .78& +11.4& -0.1 \\
\textbf{A+V+T}& 228.9& -1.2& +8.7& 205.9& -8.4& +0.2& 2.06& -12.6& +0.6& 2.08& -22.9& +0.3& .264& -0.0& +2.7& .192& -7.5& +0.5& 1.10& +0.9& +4.3& .60& -16.6& +8.4 \\

\bottomrule
\end{tabular}
}
\label{tab:regression}
\end{table*}

\textbf{Likes per day: }As with Vp/d, we see that arousal and valence are strong as singular signals when utilising all summary features; however, in this case, there is no improvement found through the fusion of multiple feature types. Further to this, the cross-task selection method appears to improve results across all types, aside from the fusion of arousal, valence, and trustworthiness. As with Vp/d, valence again obtains our best result, improved even further by the automatic selection, up to 1.23 MAE Vp/d. Although the automatic selection appears valid for valence, this was not consistent across all the feature type variations. Trustworthiness appears much weaker than all other features types in this case, although when observing scores on the development set; we see that trustworthiness is our strongest singular signal (2.21), even showing promise when fused with the other feature types and from the automatic feature selection.

\textbf{Comments per day: }Results obtained from Cp/d continue to show the trend of valence being a meaningful signal. Again for all features, as singular signal both arousal and valence show the best score (0.154 MAE for both). Valence improves by the auto-selection process, and performs better with the cross-task method. Fusion in this case generally does not show much benefit assigned from the combination of arousal and valence, in which our best Cp/d score is obtained from cross-task selection of 0.145 MAE. As previously, trustworthiness is again not the strongest signal on test, however, we see a similar strength on development set.

\textbf{Likes of comments per day: }Arousal achieves the strongest result from all features for CLp/d. Unlike the other user engagement criteria, we see a large decrease across most results from the both selection methods. The best improvement comes from the fusion of arousal and valence with the task specific selection method. However, from automatic selection, there is a large decrease. 
As in other criteria, trustworthiness again performs better than other signals on development, and poorly on test, although the cross-task selection does show improvement for trustworthiness on test, but the absolute value still does not beat that of the arousal and valence.  

\textbf{Result Discussion: }When evaluating all results across each user-engagement criteria, it appears that our cross-task feature selection approach obtains the best results more consistently than either automatic selection or all features indicating that a more general selection stabilises generalisation. Through these feature selection approaches valence appears to be a more meaningful signal for most criteria, which can be expected given the positive:negative relationship that is inherent to all the criteria. Furthermore, without any selection, arousal is clearly a strong signal for prediction: with fusion of arousal and valence for Vp/d there is also an improvement. To this end, fusion in general does in no case obtain sustainable better results. With this in mind, further fusion strategies incorporating multiple modes at various stages in the network may be beneficial for further study.   

Trustworthiness is consistently behind arousal and valence for all criteria. A somewhat unexpected result, although this may be caused by generalizability issues on the testing set, further shown by the strong results during development. Interestingly, as a single signal trustworthiness performs better than arousal and valence without feature selection for Vp/d. This result is promising, as it shows a tendency that trust is generally valuable for viewership, a finding which is supported by the literature in regards to building a parasocial relationship~\cite{lim2020role}.

\vspace{-0.2cm}
\subsection{General Discussion}
When observing the literature concerning user engagement and the potential advantage of performing this automatically - we see that one essential aspect is the ability for a content creator to develop the parasocial relationship with their viewers \cite{chapple2017investigation}. In this regard, we see that the features from each emotional dimension (arousal, valence and trustworthiness) can predict core user engagement criteria. Most notably, as we mention previously short-term \textbf{fluctuations in arousal} appear to increase user engagement, and therefore it could be assumed such emotional understanding of video content will lead to higher user-engagement (\ie an improved\textbf{ parasocial relationship}). 

Furthermore, the YouTube algorithm itself is known to bias content which has higher user engagement criterion, \eg comments and likes per day. With this in mind, integration of the emotional features identified herein (which could be utilised for predicting forthcoming user engagement, \cf \Cref{sec:future}) may result in higher user engagement in other areas, \eg views per days, resulting in better financial outcomes for the creator. The correlations between these aspects, \ie the increase of comments per day, vs views per day should be further researched concerning these emotional dimensions. 

We had expected trustworthiness to be useful for predicting user-engagement, given the aforementioned parasocial relationship theory. The results are promising for the prediction of trustworthiness. However, this does not appear to be as successful as the more conventional arousal valence emotional dimensions. The current study implements an arguably conventional method for prediction task and is limited by the data domain. Applying the trustworthiness dimension to other datasets of different domains (perhaps more popular topics, such as comedy or infotainment) where similar metadata is available may show to be more fruitful for exploring the link of trust and improved user engagement.  


\vspace{-0.2cm}
\section{LIMITATIONS AND FUTURE WORK}\label{sec:future}
In this section, we would like to point out some aspects of our work that need further exploration given the novelty of the proposed idea to use continuous emotion signals for modelling explicit user engagement. 

As with \muse, some previously collected datasets harvested YouTube as their primary source \cite{zadeh2016mosi, wollmer2013youtube}. However, they either do not provide continuous emotion signals or the video \textbf{metadata} (\eg unique video identifiers) of these datasets. Therefore, \muse is currently the only dataset that allows studies similar to this, limits extensive exploration in other domains. We want to encourage future dataset creators using social media to provide such identifiers.  

When choosing the \textbf{prediction method}, we had to make the difficult choice between interpretability and accuracy. For this study, we opted to use SVMs because we believe that initially, conceivable interactions matter more than a highly optimised outcome. This way, we can reason about relationships between influencing variables and the output predictions and compare them to ones, extracted from potential other datasets in the future. We are fully aware that state-of-the-art black-box methods, \eg deep learning, may achieve better results but lack in clarity around inner workings and may rely on spurious and non-causal correlations that are less generalisable. However, this does not mean there are no other high non-linearity interactions between inputs, which we want to explore in future work.

Another point for future exploration is the \textbf{emotional spectrum}. Although \muse provides arousal and valence, which are the most consistently used dimensions in previous research, also other third focus dimensions, for example, dominance~\cite{grimm2008vera} and likeability \cite{kossaifi2019sewa} have previously been annotated. Another interesting aspect might be categorical ratings which summarise an entire video. However, we expect much lower predictive value because of the highly compressed representation of such categories summarising the emotional content (one value instead of several dynamically extracted features based on a video-length signal). 

So far, no link existed between the use of emotional signals and user engagement. That is why, the aim of our paper was to provide a proof of concept that it is valuable to leverage such signals. However, utilising human annotations can only be the first step since they are very limited in \textbf{scalability}. The annotations are usually the prediction target for developing robust emotion recognition models. Our final process is intended to be twofold:  \begin{inparaenum}[i)]\item using audio-visual features to learn to predict the human emotional signals \item using the predicted emotional signals on unseen, unlabelled videos to extract our feature set and predict user-engagement.\end{inparaenum}\, i) is very well researched in the field achieving CCCs of more than 0.7 (high correlation between predicted and human emotional annotations) on similar data sets \cite{huang2020multimodal}. By using human annotations, we aimed to demonstrate the relationship in a vanilla way (using the targets) to avoid wrong conclusions based on any introduced prediction error bias. We also plan to explore ii) in-depth in the near future.


Through a bridge of emotion recognition and user engagement, we see novel \textbf{applications}. 
The link between emotional and user engagement provides information about what and when (e.g. a part of a video with many arousal peaks) exactly causes a user to feel e.g. aversion, interest or frustration \cite{picard1999affective}. Two parties may particularly benefit from these findings: a) Social media network providers: The relationships discovered are directly related to the user retention (e.g. user churn rate) \cite{lebreton2020predicting} and activity (e.g. recommender systems) \cite{7423759}. These are the most common and important tasks of these platforms and are still extremely difficult to model to this day \cite{liu2019characterizing, yang2018know, lin2018ll}. 
Maybe more importantly, critical, emotionally charged videos (e.g. misinformation, fake messages, hate speech) can be recognised and recommendation systems adapted accordingly.
b) Content creators (marketing, advertising): Companies act as (video) creators to interact with customers. In our work we focused to show a connection between generalizable emotional characteristics and user engagement. However, we believe that there are various weaker/stronger influenced subgroups. A company can identify and target such groups or even explicitly fine-tune their content. 

\vspace{-0.2cm}
\section{CONCLUSION}
For the first time, we have empirically (and on a large-scale) presented in this contribution that there are both, intuitive and complex relationships between user engagement indicators and continuously annotated emotion as well as trustworthiness signals in user-generated data. Of prominence, our contribution finds that emotion increases engagement when arousal is consistently bounded. 
In other words, the more consistent the portrayed arousal throughout a video, the better the engagement with it. This finding contradicted previous emotion literature~\cite{Sagha17-PTP}.
Arousal shows consistently more robust prediction results, although valence innately (given the link of positive and negative) appears to be more valuable for prediction of video likes.

Further to this, we introduce trustworthiness as a continuous `emotion' dimension for engagement, and find when utilising this for prediction, there is an overall value for monitoring user-engagement in social-media content. However, when fusing the signals, their appears to be little benefit from the current recognition paradigm. Furthermore, we assume that too strict feature selection causes generalisation issues since often promising results on the development set seem nontransferable to the test set.

From the strong correlation of the results for trustworthiness, we consider that the addition of this dimension is of use for user engagement; however, further investigation in other domains would be valuable. When applying these metrics in a cross-modal sentiment paradigm, there may also be benefits for the prediction of audio-visual hate speech likelihood, as well as fake news.


\begin{table*}[!htb]
\centering
\caption{APPENDIX: Pearson correlation matrix for arousal features}
\resizebox{1.0\linewidth}{!}{
\begin{tabular}{l||rrrrrrrrrrrrrrrrrrrrrrrr}
\toprule
index & std & q\_0.05 &q\_0.25 &median &q\_0.75 &q\_0.95 &absE & aSoC &MaCh & MCh & MsdC &CrM & peaks & skewness & kurtosis & LSBMe &LSAMe &CBMe & FiLMa & FiLMi & LaLMa & LaLMi & PReDa & SaEn \\
\midrule
 views pd &$-0.293^1$ &$0.231^1$ & $0.002$ &$-0.089$ &$-0.152^2$ &$-0.212^1$ & $0.226^1$ &$0.326^1$ &$-0.015$ &$-0.004$ & $-0.007$ &$0.343^1$ & $0.440^1$ &$0.124^2$ &$0.051$ & $0.088$ &$0.045$ & $0.461^1$ &$-0.027$ & $0.030$ &$-0.028$ & $0.029$ & $0.223^1$ &$0.067$ \\
 likes pd &$-0.268^1$ &$0.203^1$ &$-0.011$ &$-0.097$ &$-0.146^2$ &$-0.205^1$ & $0.240^1$ &$0.300^1$ &$-0.029$ &$-0.006$ &$0.001$ &$0.315^1$ & $0.434^1$ &$0.123^2$ &$0.040$ & $0.085$ &$0.046$ & $0.456^1$ &$-0.033$ & $0.028$ &$-0.033$ & $0.027$ & $0.230^1$ &$0.046$ \\
dislikes pd &$-0.245^1$ &$0.220^1$ & $0.002$ &$-0.071$ &$-0.112^3$ &$-0.166^1$ & $0.214^1$ &$0.278^1$ &$-0.020$ &$-0.016$ &$0.009$ &$0.304^1$ & $0.411^1$ &$0.121^3$ &$0.026$ & $0.103^3$ &$0.063$ & $0.428^1$ &$-0.032$ & $0.051$ &$-0.032$ & $0.050$ & $0.196^1$ &$0.055$ \\
 - like dislike ratio &$-0.087$ &$0.057$ & $0.010$ &$-0.103^3$ &$-0.076$ &$-0.020$ &$-0.066$ & $-0.016$ &$-0.003$ & $0.056$ & $-0.021$ &$0.008$ &$-0.026$ &$0.047$ &$0.063$ &$-0.036$ & $-0.037$ & $0.007$ & $0.048$ &$-0.061$ & $0.048$ &$-0.061$ & $0.049$ &$0.006$ \\
comments pd &$-0.240^1$ &$0.204^1$ & $0.004$ &$-0.075$ &$-0.112^3$ &$-0.169^1$ & $0.222^1$ &$0.280^1$ &$-0.021$ &$-0.007$ &$0.004$ &$0.297^1$ & $0.409^1$ &$0.109^3$ &$0.028$ & $0.083$ &$0.037$ & $0.424^1$ &$-0.035$ & $0.048$ &$-0.035$ & $0.047$ & $0.193^1$ &$0.052$ \\
- positive pd &$-0.240^1$ &$0.214^1$ & $0.012$ &$-0.068$ &$-0.101$ &$-0.156^2$ & $0.226^1$ &$0.286^1$ &$-0.020$ &$-0.009$ &$0.005$ &$0.308^1$ & $0.418^1$ &$0.107^3$ &$0.028$ & $0.082$ &$0.033$ & $0.428^1$ &$-0.026$ & $0.050$ &$-0.027$ & $0.049$ & $0.191^1$ &$0.056$ \\
 - positive ratio & $0.054$ & $-0.064$ &$-0.098$ &$-0.053$ &$-0.052$ & $0.019$ & $0.202$ &$0.036$ &$-0.050$ &$-0.145$ &$0.045$ &$0.026$ & $0.112$ &$0.049$ & $-0.037$ & $0.145$ &$0.087$ & $0.123$ &$-0.033$ & $0.131$ &$-0.033$ & $0.131$ & $0.043$ & $-0.030$ \\
 - neutral pd &$-0.225^1$ &$0.196^1$ & $0.009$ &$-0.067$ &$-0.102$ &$-0.151^2$ & $0.178^1$ &$0.244^1$ &$-0.020$ &$-0.003$ &$0.006$ &$0.267^1$ & $0.358^1$ &$0.101$ &$0.030$ & $0.073$ &$0.027$ & $0.374^1$ &$-0.043$ & $0.051$ &$-0.044$ & $0.050$ & $0.170^1$ &$0.047$ \\
- neutral ratio &$-0.182^2$ &$0.143^2$ & $0.037$ &$-0.036$ &$-0.109$ &$-0.078$ & $0.109^3$ &$0.050$ &$-0.150^3$ & $0.063^3$ & $-0.035$ &$0.074$ & $0.169^2$ &$0.064$ &$0.077$ & $0.112$ &$0.163$ & $0.191^1$ &$-0.044$ & $0.015$ &$-0.044$ & $0.015$ & $0.205^1$ & $-0.106$ \\
- negative pd &$-0.248^1$ &$0.191^1$ &$-0.014$ &$-0.090$ &$-0.131^2$ &$-0.198^1$ & $0.269^1$ &$0.309^1$ &$-0.022$ &$-0.009$ &$0.001$ &$0.312^1$ & $0.451^1$ &$0.116^3$ &$0.024$ & $0.093$ &$0.056$ & $0.469^1$ &$-0.029$ & $0.038$ &$-0.030$ & $0.037$ & $0.216^1$ &$0.051$ \\
 - negative ratio & $0.085$ & $-0.045$ & $0.066$ & $0.077$ & $0.130^3$ & $0.040$ &$-0.273^1$ & $-0.072$ & $0.159$ & $0.091$ & $-0.017$ & $-0.079$ &$-0.232^1$ & $-0.094$ & $-0.022$ &$-0.221^2$ & $-0.204$ &$-0.259^1$ & $0.064$ &$-0.136^3$ & $0.065^3$ &$-0.136^3$ &$-0.194^2$ &$0.108$ \\
- pos.-neg. ratio &$-0.037$ &$0.014$ &$-0.068$ &$-0.075$ &$-0.069$ &$-0.039$ & $0.131^2$ &$0.081$ &$-0.026$ &$-0.083$ &$0.048$ &$0.069$ & $0.125^3$ &$0.054$ &$0.007$ & $0.095^3$ &$0.034$ & $0.137^2$ &$-0.074$ & $0.127^2$ &$-0.074$ & $0.127^2$ & $0.062$ &$0.003$ \\
 likes of comments pd &$-0.249^1$ &$0.206^1$ &$-0.007$ &$-0.088$ &$-0.125^2$ &$-0.186^1$ & $0.223^1$ &$0.292^1$ &$-0.027$ &$-0.008$ &$0.006$ &$0.309^1$ & $0.414^1$ &$0.120^3$ &$0.037$ & $0.088$ &$0.047$ & $0.441^1$ &$-0.009$ & $0.003$ &$-0.010$ & $0.002$ & $0.222^1$ &$0.041$ \\
 - positive com. pd &$-0.231^1$ &$0.214^1$ & $0.026$ &$-0.049$ &$-0.084$ &$-0.139^2$ & $0.187^1$ &$0.305^1$ &$-0.002$ &$-0.009$ & $-0.002$ &$0.331^1$ & $0.391^1$ &$0.084$ &$0.041$ & $0.050$ &$0.009$ & $0.390^1$ &$-0.015$ & $0.020$ &$-0.015$ & $0.019$ & $0.181^1$ &$0.068$ \\
- positive com. ratio &$-0.045$ & $-0.026$ &$-0.028$ &$-0.023$ &$-0.068$ &$-0.055$ & $0.122$ &$0.056$ &$-0.000$ &$-0.012$ &$0.016$ &$0.074$ & $0.105$ &$0.004$ &$0.042$ & $0.047$ &$0.014$ & $0.110$ &$-0.039$ &$-0.010$ &$-0.039$ &$-0.010$ & $0.034$ &$0.019$ \\
- neutral com. pd &$-0.228^1$ &$0.188^1$ &$-0.016$ &$-0.094$ &$-0.128^2$ &$-0.170^1$ & $0.196^1$ &$0.242^1$ &$-0.038$ &$-0.007$ &$0.015$ &$0.261^1$ & $0.368^1$ &$0.129^2$ &$0.034$ & $0.075$ &$0.043$ & $0.413^1$ & $0.009$ &$-0.000$ & $0.009$ &$-0.001$ & $0.216^1$ &$0.020$ \\
 - neutral com. ratio &$-0.059$ &$0.087$ & $0.049$ & $0.002$ &$-0.015$ & $0.047$ & $0.059$ &$0.021$ &$-0.112$ &$-0.104^3$ &$0.127^3$ &$0.028$ & $0.080$ &$0.045$ &$0.061$ & $0.035$ &$0.026$ & $0.096$ &$-0.044$ & $0.101$ &$-0.044$ & $0.100$ & $0.147^2$ & $-0.079$ \\
 - negative com. pd &$-0.232^1$ &$0.152^2$ &$-0.048$ &$-0.120^3$ &$-0.153^2$ &$-0.233^1$ & $0.255^1$ &$0.249^1$ &$-0.045$ &$-0.003$ &$0.005$ &$0.245^1$ & $0.393^1$ &$0.131^2$ &$0.021$ & $0.147^2$ &$0.105^3$ & $0.432^1$ &$-0.026$ &$-0.021$ &$-0.026$ &$-0.022$ & $0.230^1$ &$0.012$ \\
- negative com. ratio & $0.089^3$ & $-0.037$ &$-0.007$ & $0.022$ & $0.081$ & $0.022$ &$-0.168^3$ & $-0.073$ & $0.082$ & $0.089$ & $-0.109$ & $-0.096$ &$-0.165^2$ & $-0.038$ & $-0.087$ &$-0.073$ & $-0.033$ &$-0.182^2$ & $0.072^3$ &$-0.064$ & $0.072^3$ &$-0.064$ &$-0.142^3$ &$0.038$ \\
 like-comment ratio &$-0.271^1$ &$0.248^1$ &$-0.010$ &$-0.091$ &$-0.164^2$ &$-0.182^2$ & $0.150^3$ &$0.139^3$ &$-0.087$ &$-0.065$ &$0.058$ &$0.161^2$ & $0.257^1$ &$0.155^2$ &$0.073$ & $0.154^2$ &$0.090$ & $0.282^1$ &$-0.017$ &$-0.095^3$ &$-0.018$ &$-0.097^3$ & $0.097$ &$0.006$ \\
 - positive lc. ratio &$-0.288^1$ &$0.154^2$ &$-0.033$ &$-0.117^3$ &$-0.199^1$ &$-0.254^1$ & $0.257^1$ &$0.308^1$ &$-0.021$ &$-0.058$ &$0.026$ &$0.311^1$ & $0.444^1$ &$0.120^3$ &$0.067$ & $0.073$ &$0.035$ & $0.441^1$ &$-0.051$ & $0.018$ &$-0.052$ & $0.016$ & $0.187^1$ &$0.069$ \\
 - negative lc. ratio &$-0.220^1$ &$0.107$ &$-0.050$ &$-0.105$ &$-0.155^2$ &$-0.231^1$ & $0.286^1$ &$0.201^1$ &$-0.069$ &$-0.075$ &$0.056$ &$0.195^1$ & $0.374^1$ &$0.118^3$ &$0.020$ & $0.181^1$ &$0.166^2$ & $0.398^1$ &$-0.055$ & $0.039$ &$-0.056$ & $0.038$ & $0.215^1$ & $-0.008$ \\
 mean &$-0.166^1$ &$0.130^1$ &$-0.008^1$ &$-0.064^1$ &$-0.093^1$ &$-0.123^1$ & $0.148^1$ &$0.175^1$ &$-0.024^1$ &$-0.015^1$ &$0.011^1$ &$0.185^1$ & $0.260^1$ &$0.081^1$ &$0.028^1$ & $0.067^1$ &$0.039^1$ & $0.275^1$ &$-0.020^1$ & $0.017^1$ &$-0.020^1$ & $0.016^1$ & $0.135^1$ &$0.025^1$ \\
\bottomrule
\end{tabular}
}
\label{tab:pval_arousal}
\end{table*}

\begin{table*}[!htb]
\centering
\caption{APPENDIX: Pearson correlation matrix for valence features.}
\resizebox{1.0\linewidth}{!}{
\begin{tabular}{l||rrrrrrrrrrrrrrrrrrrrrrrr}
\toprule
 & std & q\_0.05 &q\_0.25 & median &q\_0.75 &q\_0.95 & absE &aSoC &MaCh & MCh &MsdC & CrM & peaks &skewness &kurtosis & LSBMe & LSAMe &CBMe & FiLMa &FiLMi & LaLMa &LaLMi & PReDa &SaEn \\
\midrule
 views pd & $0.038$ & $-0.031$ & $0.133^2$ &$0.173^1$ & $0.139^2$ & $0.053$ &$0.469^1$ & $0.042$ &$-0.160^1$ &$-0.069$ & $0.071$ &$-0.023$ & $0.381^1$ &$-0.163^1$ & $0.032$ & $0.431^1$ & $0.308^1$ & $0.375^1$ &$-0.039$ &$0.035$ &$-0.039$ &$0.031$ & $0.182^1$ &$-0.176^1$ \\
 likes pd &$-0.006$ &$0.004$ & $0.115^3$ &$0.139^2$ & $0.100^3$ & $0.015$ &$0.424^1$ & $0.070$ &$-0.138^2$ &$-0.066$ & $0.062$ & $0.017$ & $0.385^1$ &$-0.132^2$ & $0.034$ & $0.410^1$ & $0.282^1$ & $0.382^1$ &$-0.032$ &$0.018$ &$-0.032$ &$0.014$ & $0.179^1$ &$-0.143^2$ \\
dislikes pd &$-0.039$ &$0.031$ & $0.082$ &$0.098$ & $0.063$ &$-0.022$ &$0.356^1$ & $0.044$ &$-0.146^2$ &$-0.076$ & $0.069$ & $0.008$ & $0.347^1$ &$-0.098$ & $0.027$ & $0.380^1$ & $0.271^1$ & $0.367^1$ &$-0.006$ &$0.013$ &$-0.006$ &$0.008$ & $0.182^1$ &$-0.148^1$ \\
 - like dislike ratio &$-0.011$ &$0.004$ & $0.064$ &$0.073$ & $0.022$ & $0.028$ &$0.019$ &$-0.083$ &$-0.065$ & $0.048$ &$-0.052$ &$-0.106^3$ &$-0.029$ &$-0.060$ & $0.039$ &$-0.127^2$ & $0.042$ &$-0.020$ &$-0.020$ &$0.001$ &$-0.020$ &$0.002$ & $0.043$ &$-0.082$ \\
comments pd &$-0.036$ &$0.029$ & $0.082$ &$0.097$ & $0.062$ &$-0.017$ &$0.352^1$ & $0.050$ &$-0.140^2$ &$-0.063$ & $0.053$ & $0.020$ & $0.344^1$ &$-0.095$ & $0.031$ & $0.370^1$ & $0.270^1$ & $0.366^1$ &$-0.002$ &$0.011$ &$-0.002$ &$0.007$ & $0.185^1$ &$-0.146^2$ \\
- positive pd &$-0.063$ &$0.043$ & $0.092$ &$0.094$ & $0.052$ &$-0.033$ &$0.340^1$ & $0.059$ &$-0.136^2$ &$-0.066$ & $0.056$ & $0.034$ & $0.353^1$ &$-0.101$ & $0.048$ & $0.363^1$ & $0.255^1$ & $0.372^1$ & $0.023$ &$0.010$ & $0.023$ &$0.006$ & $0.193^1$ &$-0.138^2$ \\
 - positive ratio &$-0.170^1$ &$0.051^2$ & $0.043^2$ & $-0.035$ &$-0.110^2$ &$-0.149^1$ &$0.041$ & $0.123^3$ &$-0.008$ &$-0.123$ & $0.013$ & $0.154^1$ & $0.157$ &$-0.068^3$ & $0.135^1$ & $0.029^3$ & $0.039^3$ & $0.135$ & $0.008$ & $-0.065$ & $0.007$ & $-0.057$ & $0.095$ & $0.080^2$ \\
 - neutral pd &$-0.034$ &$0.026$ & $0.062$ &$0.077$ & $0.047$ &$-0.028$ &$0.301^1$ & $0.019$ &$-0.134^2$ &$-0.055$ & $0.047$ &$-0.000$ & $0.293^1$ &$-0.079$ & $0.027$ & $0.340^1$ & $0.256^1$ & $0.319^1$ &$-0.004$ &$0.005$ &$-0.004$ &$0.001$ & $0.170^1$ &$-0.144^2$ \\
- neutral ratio &$-0.025$ & $-0.058$ & $0.038$ &$0.033$ &$-0.013$ &$-0.053$ &$0.204^1$ & $0.053$ &$-0.013$ &$-0.026$ & $0.039$ & $0.031$ & $0.170^2$ &$-0.075$ & $0.087$ & $0.236^1$ & $0.240^1$ & $0.170^2$ &$-0.008$ &$0.009$ &$-0.008$ &$0.001$ & $0.147^1$ &$-0.072^2$ \\
- negative pd &$-0.008$ &$0.018$ & $0.095$ &$0.124^2$ & $0.092^3$ & $0.017$ &$0.416^1$ & $0.084$ &$-0.143^2$ &$-0.067$ & $0.054$ & $0.034$ & $0.392^1$ &$-0.106^3$ & $0.016$ & $0.401^1$ & $0.288^1$ & $0.406^1$ &$-0.026$ &$0.022$ &$-0.026$ &$0.018$ & $0.190^1$ &$-0.147^2$ \\
 - negative ratio & $0.180^1$ & $-0.005$ &$-0.069^2$ &$0.009$ & $0.115$ & $0.181^1$ & $-0.190$ &$-0.156^3$ & $0.017$ & $0.137$ &$-0.042$ &$-0.169^1$ &$-0.276^1$ & $0.121^1$ &$-0.194^1$ &$-0.204$ &$-0.216^2$ &$-0.255^1$ &$-0.001$ &$0.055$ &$-0.000$ &$0.054$ &$-0.200^1$ &$-0.022$ \\
- pos.-neg. ratio &$-0.276^1$ &$0.110^2$ & $0.171^1$ &$0.027$ &$-0.090$ &$-0.152^2$ &$0.034$ & $0.115^3$ & $0.019$ &$-0.087$ & $0.016$ & $0.121^3$ & $0.152^2$ &$-0.180^1$ & $0.313^1$ &$-0.050$ &$-0.029$ & $0.126^3$ &$-0.053$ & $-0.087$ &$-0.053$ & $-0.087$ & $0.099^3$ & $0.116^3$ \\
 likes of comments pd & $0.000$ &$0.012$ & $0.108^3$ &$0.138^2$ & $0.108^3$ & $0.032$ &$0.407^1$ & $0.051$ &$-0.139^2$ &$-0.064$ & $0.054$ & $0.002$ & $0.367^1$ &$-0.119^3$ & $0.021$ & $0.394^1$ & $0.263^1$ & $0.366^1$ &$-0.010$ &$0.021$ &$-0.010$ &$0.017$ & $0.156^1$ &$-0.147^2$ \\
 - positive com. pd &$-0.004$ & $-0.013$ & $0.116^3$ &$0.130^2$ & $0.094^3$ & $0.012$ &$0.360^1$ & $0.041$ &$-0.125^2$ &$-0.063$ & $0.050$ & $0.005$ & $0.326^1$ &$-0.136^2$ & $0.046$ & $0.366^1$ & $0.249^1$ & $0.316^1$ & $0.039$ &$0.029$ & $0.039$ &$0.025$ & $0.145^2$ &$-0.134^2$ \\
- positive com. ratio &$-0.082^3$ & $-0.013$ & $0.114^2$ &$0.041$ & $0.022$ &$-0.079$ &$0.081$ & $0.090$ & $0.006$ &$-0.028$ & $0.063$ & $0.119^2$ & $0.127^3$ &$-0.133^2$ & $0.115^2$ & $0.049$ & $0.142$ & $0.101$ &$-0.039$ & $-0.077$ &$-0.039$ & $-0.079$ & $0.077$ & $0.053$ \\
- neutral com. pd &$-0.006$ &$0.034$ & $0.065$ &$0.103^3$ & $0.084$ & $0.029$ &$0.362^1$ & $0.040$ &$-0.126^2$ &$-0.054$ & $0.051$ &$-0.002$ & $0.332^1$ &$-0.070$ &$-0.009$ & $0.356^1$ & $0.224^1$ & $0.351^1$ &$-0.035$ & $-0.008$ &$-0.036$ & $-0.012$ & $0.141^2$ &$-0.134^2$ \\
 - neutral com. ratio &$-0.046$ & $-0.006$ &$-0.005$ & $-0.011$ &$-0.039$ &$-0.059$ &$0.069$ & $0.073$ & $0.059$ &$-0.089$ &$-0.121^3$ & $0.071$ & $0.104$ &$-0.009$ & $0.054$ & $0.093^3$ & $0.000$ & $0.090$ &$-0.028$ &$0.033$ &$-0.028$ &$0.035$ & $0.056$ & $0.020$ \\
 - negative com. pd & $0.017$ &$0.016$ & $0.123^2$ &$0.159^1$ & $0.135^2$ & $0.060$ &$0.426^1$ & $0.071$ &$-0.139^2$ &$-0.063$ & $0.048$ & $0.004$ & $0.373^1$ &$-0.125^2$ & $0.015$ & $0.372^1$ & $0.262^1$ & $0.359^1$ &$-0.054$ &$0.045$ &$-0.054$ &$0.041$ & $0.147^2$ &$-0.142^2$ \\
- negative com. ratio & $0.117^2$ &$0.018$ &$-0.113^2$ & $-0.034$ & $0.005$ & $0.124^2$ & $-0.133$ &$-0.145^2$ &$-0.049$ & $0.093$ & $0.024$ &$-0.173^1$ &$-0.205^1$ & $0.142^1$ &$-0.157^1$ &$-0.117$ &$-0.145$ &$-0.169^2$ & $0.060$ &$0.054$ & $0.060$ &$0.055$ &$-0.119^3$ &$-0.068$ \\
 like-comment ratio &$-0.003$ & $-0.028$ & $0.237^1$ &$0.260^1$ & $0.202^1$ & $0.043$ &$0.334^1$ & $0.080$ &$-0.121^2$ &$-0.149$ & $0.088$ &$-0.014$ & $0.251^1$ &$-0.250^1$ & $0.071$ & $0.243^1$ & $0.205^1$ & $0.212^1$ &$-0.053$ &$0.134^2$ &$-0.054$ &$0.122^2$ & $0.136^3$ &$-0.100^3$ \\
 - positive lc. ratio &$-0.024$ & $-0.013$ & $0.156^2$ &$0.169^1$ & $0.109^2$ & $0.013$ &$0.430^1$ & $0.157^2$ &$-0.098$ &$-0.097$ & $0.075$ & $0.093$ & $0.392^1$ &$-0.182^1$ & $0.088$ & $0.366^1$ & $0.249^1$ & $0.380^1$ &$-0.091$ &$0.019$ &$-0.091$ &$0.011$ & $0.192^1$ &$-0.077$ \\
 - negative lc. ratio &$-0.045$ &$0.051$ & $0.129^2$ &$0.161^1$ & $0.110^2$ & $0.007$ &$0.400^1$ & $0.164^2$ &$-0.100$ &$-0.129$ & $0.086$ & $0.085$ & $0.383^1$ &$-0.154^2$ & $0.082$ & $0.326^1$ & $0.208^1$ & $0.360^1$ &$-0.085$ &$0.027$ &$-0.086$ &$0.019$ & $0.169^2$ &$-0.066$ \\
 mean &$-0.024^1$ &$0.013^1$ & $0.084^1$ &$0.092^1$ & $0.059^1$ & $0.001^1$ &$0.250^1$ & $0.047^1$ &$-0.085^1$ &$-0.053^1$ & $0.037^1$ & $0.014^1$ & $0.233^1$ &$-0.094^1$ & $0.042^1$ & $0.228^1$ & $0.167^1$ & $0.232^1$ &$-0.021^1$ &$0.014^1$ &$-0.021^1$ &$0.010^1$ & $0.117^1$ &$-0.083^1$ \\
\bottomrule
\end{tabular}
}
\label{tab:valence_ext}
\end{table*}

\begin{table*}[!htb]
\centering

\caption{APPENDIX: Pearson correlation matrix for trustworthiness features.}
\resizebox{1.0\linewidth}{!}{
\begin{tabular}{l||rrrrrrrrrrrrrrrrrrrrrrrr}
\toprule
index & std &$q_{.05}$ & $q_{.25}$ & median & $q_{.75}$ & $q_{.95}$ &absE &aSoC &MaCh & MCh & MsdC & CrM & peaks &skew & kurtosis &LSBMe & LSAMe &CBMe & FiLMa &FiLMi & LaLMa &LaLMi & PReDa &SaEn \\
\midrule
 views pd &$-0.305^1$ & $0.356^1$ &$0.295^1$ &$0.229^1$ &$0.175^1$ &$0.116^3$ & $0.496^1$ & $0.149^2$ &$-0.117^3$ &$-0.098$ &$0.027$ &$-0.076$ & $0.423^1$ &$-0.368^1$ &$0.410^1$ &$0.045$ & $0.195^1$ & $0.326^1$ &$-0.080$ &$0.115$ &$-0.081$ &$0.114$ & $0.344^1$ &$-0.012$ \\
 likes pd &$-0.288^1$ & $0.335^1$ &$0.272^1$ &$0.205^1$ &$0.152^2$ &$0.103^3$ & $0.473^1$ & $0.139^2$ &$-0.122^2$ &$-0.088$ &$0.019$ &$-0.069$ & $0.415^1$ &$-0.349^1$ &$0.400^1$ &$0.067$ & $0.192^1$ & $0.333^1$ &$-0.050$ &$0.110$ &$-0.051$ &$0.110$ & $0.348^1$ &$-0.022$ \\
dislikes pd &$-0.275^1$ & $0.306^1$ &$0.262^1$ &$0.197^1$ &$0.145^2$ &$0.095$ & $0.436^1$ & $0.163^1$ &$-0.083$ &$-0.092$ &$0.020$ &$-0.052$ & $0.393^1$ &$-0.317^1$ &$0.347^1$ &$0.058$ & $0.196^1$ & $0.305^1$ &$-0.044$ &$0.123^3$ &$-0.045$ &$0.123^3$ & $0.302^1$ & $0.014$ \\
 - like dislike ratio &$-0.044$ & $0.023$ & $-0.068$ & $-0.058$ & $-0.052$ & $-0.028$ &$-0.045$ & $0.046$ & $0.066$ & $0.024$ & $-0.006$ & $0.039$ & $0.018$ & $0.063$ & $-0.031$ &$0.020$ &$-0.049$ & $0.031$ &$-0.066$ & $-0.053$ &$-0.066$ & $-0.053$ &$-0.081$ & $0.068$ \\
comments pd &$-0.255^1$ & $0.291^1$ &$0.232^1$ &$0.172^1$ &$0.125^2$ &$0.084$ & $0.422^1$ & $0.138^2$ &$-0.101$ &$-0.078$ &$0.013$ &$-0.050$ & $0.383^1$ &$-0.294^1$ &$0.336^1$ &$0.084$ & $0.194^1$ & $0.314^1$ &$-0.020$ &$0.113^3$ &$-0.021$ &$0.113^3$ & $0.304^1$ &$-0.011$ \\
- positive pd &$-0.257^1$ & $0.293^1$ &$0.234^1$ &$0.173^1$ &$0.126^2$ &$0.086$ & $0.430^1$ & $0.138^2$ &$-0.104^3$ &$-0.082$ &$0.013$ &$-0.041$ & $0.386^1$ &$-0.293^1$ &$0.339^1$ &$0.086$ & $0.212^1$ & $0.320^1$ &$-0.030$ &$0.116^3$ &$-0.031$ &$0.116^3$ & $0.307^1$ &$-0.014$ \\
 - positive ratio &$-0.014$ & $0.016$ & $-0.050$ & $-0.031$ & $-0.033$ & $-0.047$ & $0.073$ & $0.100$ &$-0.046$ &$-0.151$ &$0.081$ & $0.104$ & $0.109$ & $0.039$ & $-0.021$ &$0.138$ & $0.113$ & $0.150$ &$-0.035$ &$0.002$ &$-0.034$ &$0.002$ & $0.154^3$ &$-0.036$ \\
 - neutral pd &$-0.222^1$ & $0.261^1$ &$0.210^1$ &$0.158^1$ &$0.117^3$ &$0.080$ & $0.377^1$ & $0.131^2$ &$-0.083$ &$-0.071$ &$0.010$ &$-0.044$ & $0.340^1$ &$-0.262^1$ &$0.295^1$ &$0.067$ & $0.181^1$ & $0.264^1$ &$-0.005$ &$0.108$ &$-0.006$ &$0.107$ & $0.262^1$ &$-0.003$ \\
- neutral ratio &$-0.224^1$ & $0.200^1$ &$0.097^2$ &$0.048$ &$0.013$ & $-0.024$ & $0.150^2$ & $0.197^1$ & $0.096^2$ &$-0.078$ & $-0.042$ & $0.155^2$ & $0.222^1$ &$-0.066$ &$0.103^3$ & $-0.019$ & $0.084$ & $0.169^2$ &$-0.139^2$ &$0.146^3$ &$-0.141^2$ &$0.145^3$ & $0.039$ & $0.163^1$ \\
- negative pd &$-0.286^1$ & $0.317^1$ &$0.248^1$ &$0.181^1$ &$0.128^2$ &$0.083$ & $0.455^1$ & $0.139^2$ &$-0.120^3$ &$-0.078$ &$0.018$ &$-0.065$ & $0.422^1$ &$-0.324^1$ &$0.374^1$ &$0.103$ & $0.182^1$ & $0.362^1$ &$-0.030$ &$0.112$ &$-0.031$ &$0.112$ & $0.345^1$ &$-0.021$ \\
 - negative ratio & $0.180^2$ &$-0.165^1$ & $-0.025$ & $-0.006$ &$0.022$ &$0.063$ &$-0.182^1$ &$-0.241^1$ &$-0.027$ & $0.202$ & $-0.046$ &$-0.214^1$ &$-0.269^1$ & $0.011$ & $-0.057$ & $-0.116$ &$-0.170^2$ &$-0.268^1$ & $0.137^3$ & $-0.111$ & $0.138^3$ & $-0.110$ &$-0.176^3$ &$-0.088$ \\
- pos.-neg. ratio &$-0.120$ & $0.115$ &$0.104$ &$0.080$ &$0.060$ &$0.042$ & $0.156^2$ & $0.056$ &$-0.075$ &$-0.088$ &$0.023$ & $0.078$ & $0.132^2$ &$-0.074$ &$0.085$ &$0.067$ & $0.100^3$ & $0.112^3$ & $0.036$ &$0.049$ & $0.035$ &$0.049$ & $0.167^2$ &$-0.048$ \\
 likes of comments pd &$-0.281^1$ & $0.310^1$ &$0.247^1$ &$0.180^1$ &$0.130^2$ &$0.081$ & $0.446^1$ & $0.139^2$ &$-0.112^3$ &$-0.081$ &$0.020$ &$-0.064$ & $0.393^1$ &$-0.317^1$ &$0.372^1$ &$0.049$ & $0.156^2$ & $0.323^1$ &$-0.056$ &$0.097$ &$-0.057$ &$0.097$ & $0.320^1$ &$-0.013$ \\
 - positive com. pd &$-0.252^1$ & $0.294^1$ &$0.239^1$ &$0.177^1$ &$0.132^2$ &$0.091$ & $0.427^1$ & $0.095$ &$-0.125^2$ &$-0.080$ &$0.014$ &$-0.068$ & $0.347^1$ &$-0.302^1$ &$0.352^1$ &$0.050$ & $0.185^1$ & $0.276^1$ &$-0.058$ &$0.088$ &$-0.059$ &$0.088$ & $0.309^1$ &$-0.045$ \\
- positive com. ratio &$-0.122^2$ & $0.108^3$ & $-0.034$ & $-0.068$ & $-0.065$ & $-0.051$ & $0.058$ & $0.092$ & $0.012$ &$-0.050$ &$0.141^3$ & $0.082$ & $0.107$ & $0.012$ &$0.051$ &$0.079$ & $0.002$ & $0.141^3$ &$-0.038$ &$0.013$ &$-0.037$ &$0.012$ & $0.098$ & $0.035$ \\
- neutral com. pd &$-0.250^1$ & $0.266^1$ &$0.209^1$ &$0.152^2$ &$0.109^3$ &$0.065$ & $0.395^1$ & $0.173^1$ &$-0.069$ &$-0.067$ &$0.017$ &$-0.046$ & $0.368^1$ &$-0.269^1$ &$0.325^1$ &$0.027$ & $0.114^3$ & $0.307^1$ &$-0.067$ &$0.074$ &$-0.068$ &$0.073$ & $0.266^1$ & $0.032$ \\
 - neutral com. ratio &$-0.084$ & $0.054$ &$0.051$ &$0.024$ & $-0.005$ & $-0.039$ & $0.078$ & $0.053$ & $0.004$ &$-0.047$ &$0.007$ & $0.044$ & $0.093$ &$-0.025$ &$0.022$ &$0.017$ & $0.094$ & $0.079$ &$-0.044$ &$0.055$ &$-0.043$ &$0.055$ & $0.018$ & $0.031$ \\
 - negative com. pd &$-0.287^1$ & $0.305^1$ &$0.238^1$ &$0.170^1$ &$0.115^3$ &$0.064$ & $0.411^1$ & $0.118^3$ &$-0.118^3$ &$-0.077$ &$0.028$ &$-0.064$ & $0.384^1$ &$-0.313^1$ &$0.361^1$ &$0.064$ & $0.117^3$ & $0.325^1$ &$-0.017$ &$0.117^3$ &$-0.018$ &$0.117^3$ & $0.319^1$ &$-0.021$ \\
- negative com. ratio & $0.186^1$ &$-0.149^2$ & $-0.003$ &$0.051$ &$0.069$ &$0.081$ &$-0.116^3$ &$-0.132$ &$-0.015$ & $0.086^3$ & $-0.149$ &$-0.116$ &$-0.177^2$ & $0.007$ & $-0.068$ & $-0.093$ &$-0.071$ &$-0.202^1$ & $0.071$ & $-0.053$ & $0.069$ & $-0.052$ &$-0.114$ &$-0.059$ \\
 like-comment ratio &$-0.287^1$ & $0.293^1$ &$0.266^1$ &$0.214^1$ &$0.158^2$ &$0.068$ & $0.267^1$ & $0.170^2$ & $0.002$ &$-0.235^1$ &$0.080$ & $0.011$ & $0.259^1$ &$-0.278^1$ &$0.265^1$ & $-0.064$ & $0.094$ & $0.212^1$ &$-0.223^1$ &$0.243^1$ &$-0.225^1$ &$0.242^1$ & $0.155$ & $0.104^3$ \\
 - positive lc. ratio &$-0.299^1$ & $0.325^1$ &$0.305^1$ &$0.231^1$ &$0.172^1$ &$0.114^3$ & $0.457^1$ & $0.125^3$ &$-0.146^2$ &$-0.140^2$ &$0.031$ &$-0.058$ & $0.401^1$ &$-0.350^1$ &$0.361^1$ &$0.092$ & $0.263^1$ & $0.329^1$ &$-0.078$ &$0.166^2$ &$-0.079$ &$0.166^2$ & $0.381^1$ &$-0.049$ \\
 - negative lc. ratio &$-0.279^1$ & $0.266^1$ &$0.305^1$ &$0.233^1$ &$0.170^1$ &$0.099$ & $0.400^1$ & $0.131^3$ &$-0.117^3$ &$-0.145^2$ &$0.060$ &$-0.048$ & $0.376^1$ &$-0.326^1$ &$0.307^1$ &$0.071$ & $0.231^1$ & $0.300^1$ &$-0.020$ &$0.221^1$ &$-0.020$ &$0.221^1$ & $0.337^1$ &$-0.014$ \\
 mean &$-0.185^1$ & $0.201^1$ &$0.165^1$ &$0.123^1$ &$0.089^1$ &$0.056^1$ & $0.276^1$ & $0.096^1$ &$-0.064^1$ &$-0.069^1$ &$0.017^1$ &$-0.025^1$ & $0.251^1$ &$-0.200^1$ &$0.224^1$ &$0.040^1$ & $0.119^1$ & $0.205^1$ &$-0.039^1$ &$0.084^1$ &$-0.040^1$ &$0.084^1$ & $0.200^1$ &$-0.000^1$ \\
\bottomrule
\end{tabular}
}
\label{tab:trust_extend}
\end{table*}

\newpage
\appendix[Correlation matrix]

The tables show the full correlation matrix (\cf \Cref{sec:corr_measure}) of arousal in \Cref{tab:pval_arousal}, valence in \Cref{tab:valence_ext} and trustworthiness in \Cref{tab:trust_extend} features: standard deviation ($std$), quantile ($q_x$), absolute energy (\textit{absE}), mean absolute change (\textit{MACh}), mean change (\textit{MCh}), mean central approximation of the second derivatives (\textit{MSDC}), crossings of a point m (\textit{CrM}), peaks, skewness, kurtosis, strike above the mean (\textit{LSAMe}), strike below the mean (\textit{LSBMe}), count below mean (\textit{CBMe}), absolute sum of changes (\textit{ASOC}), first and last location of the minimum and maximum (\textit{FLMi}, \textit{LLMi}, \textit{FLMa}, \textit{FLMa}), perc. of reoccurring datapoints ($PreDa$), and sample entropy (\textit{SaEn}). Superscript indicates level of statistical significance (t) at ${^1} 0.01$, ${^2} 0.05$, and ${^3} 0.1$.

\ifCLASSOPTIONcaptionsoff
  \newpage
\fi

\newpage
\newpage
\newpage
\bibliographystyle{IEEEtran}
\section{References}
\bibliography{main}

\newpage
\begin{IEEEbiography}[{\includegraphics[width=1in,height=1.25in,clip,keepaspectratio]{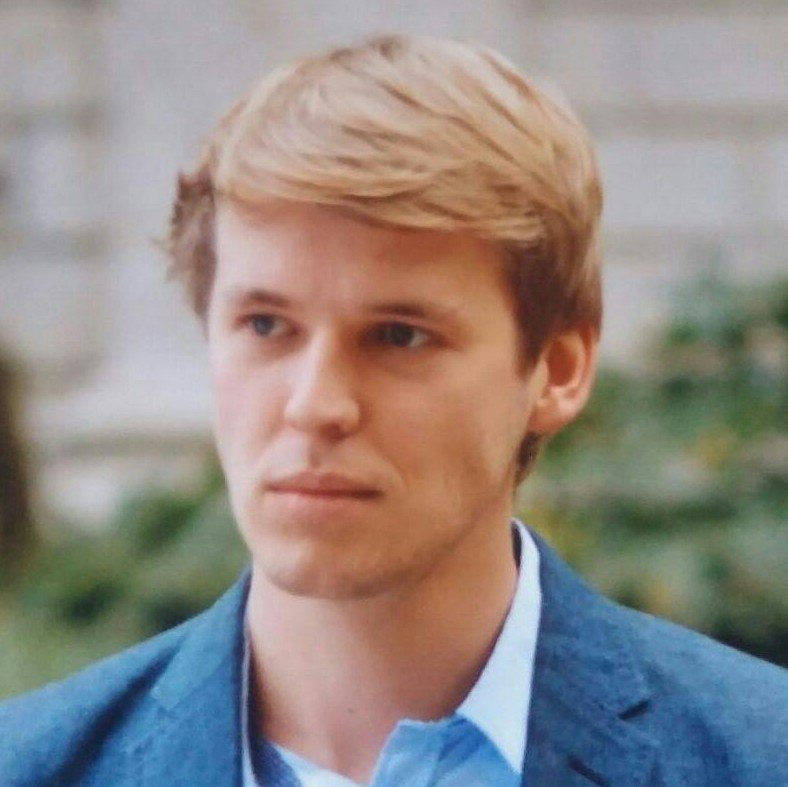}}]{Lukas Stappen}
received his Master of Science in Data Science with distinction from King's College London in 2017. He then joined
the group for Machine Learning in Health Informatics. Currently, he is a PhD candidate at the Chair for Embedded Intelligence for Health Care and Wellbeing, University of Augsburg, Germany, and a PhD Fellow of the BMW Group. His research interests include affective computing, multimodal sentiment analysis, and multimodal/cross-modal representation learning with a core focus 
on `in-the-wild' environments.
\end{IEEEbiography}

\begin{IEEEbiography}[{\includegraphics[width=1in,height=1.25in,clip,keepaspectratio]{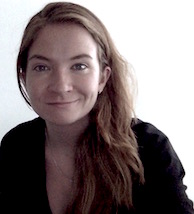}}]{Alice Baird}
received her MFA in Sound Art from Columbia University's Computer Music Center and is currently a Ph.D Fellow of the ZD.B, supervised by Professor Prof.\ Björn Schuller at the Chair of Embedded Intelligence for Healthcare and Wellbeing, University of Augsburg, Germany. Her research is focused on intelligent audio analysis in the domain of speech and general audio, and her research
interests include: health informatics, affective computing, computational paralinguistics, and
speech pathology.
\end{IEEEbiography}

\begin{IEEEbiography}[{\includegraphics[width=1in,height=1.25in,clip,keepaspectratio]{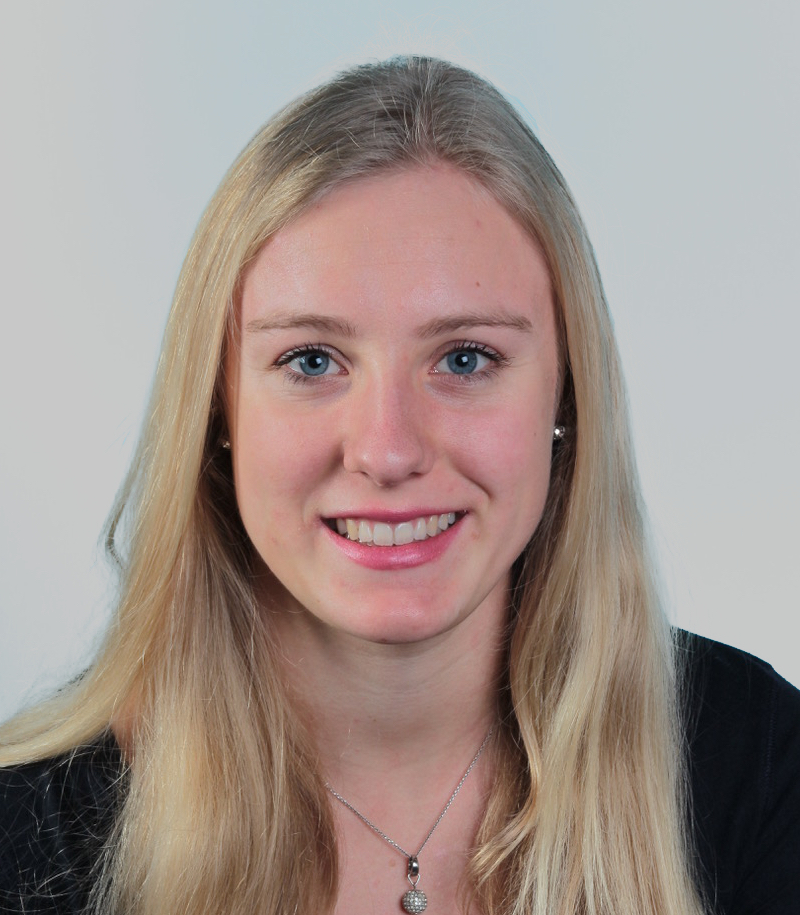}}]{Michelle Lienhart}
received her B.Sc. degree in Physics from University of Augsburg, Germany, in 2020, where
she is currently working towards her M.Sc degree in Physics and B.Sc. degree in Computer Science.
In 2019 she was a student researcher at Lawrence Berkeley National Laboratory, California, USA.
Her research interests include applications of deep learning in natural sciences and optical modulation of quantum dots for quantum technologies.
\end{IEEEbiography}

\begin{IEEEbiography}
[{\includegraphics[width=1in,height=1.25in,clip,keepaspectratio]{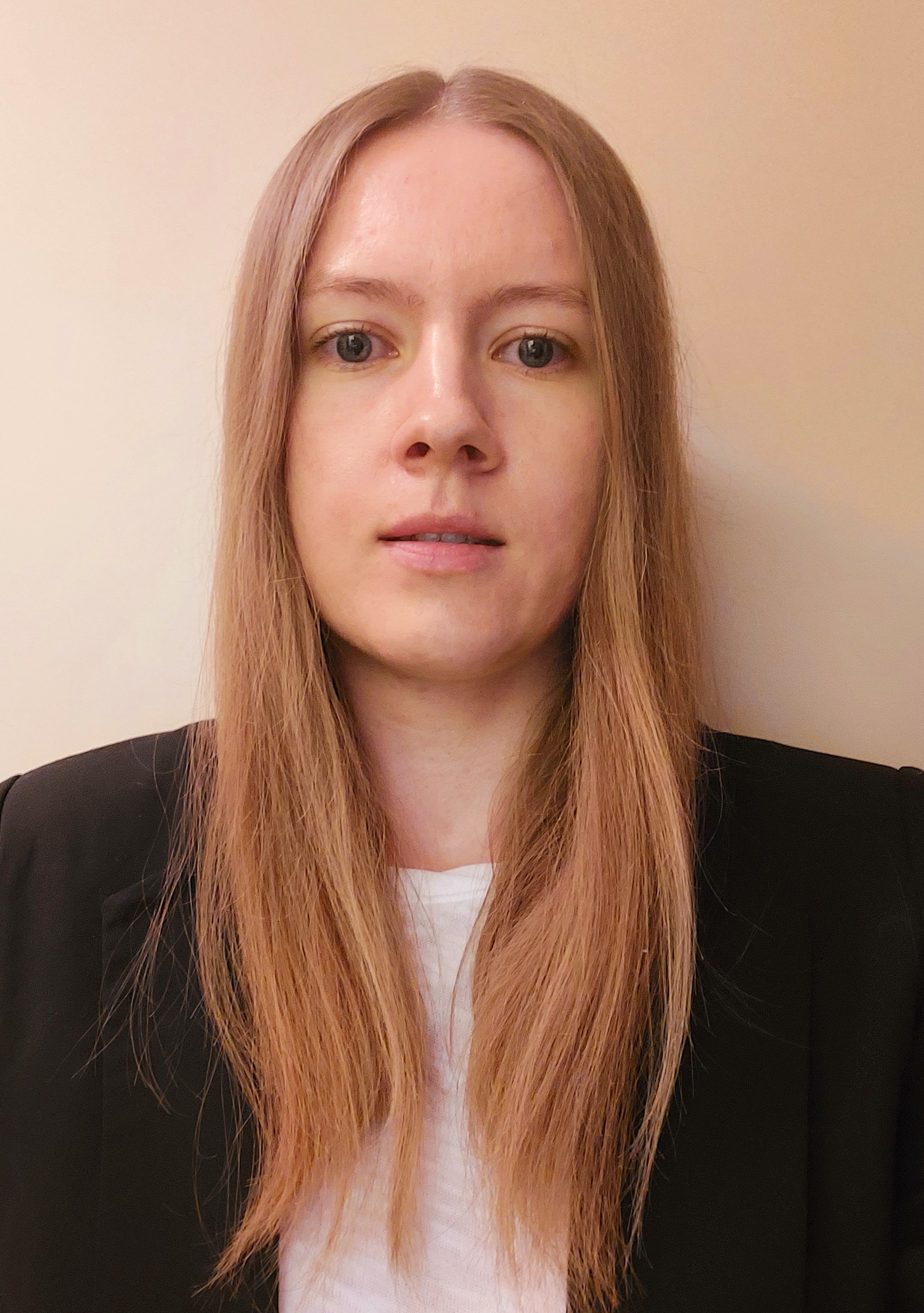}}]{Annalena Bätz}
received her Bachelor of Science degree in Business Information Systems from University of Augsburg, Germany in 2020. There, she is currently doing her Master of Science in Business Information Systems. Her research is focused on the application of AI and its business influence, Big Data Analytics, Digital Business Transformation and Business Process Management.
\end{IEEEbiography}

\begin{IEEEbiography}
    [{\includegraphics[width=1in,height=1.25in,clip,keepaspectratio]{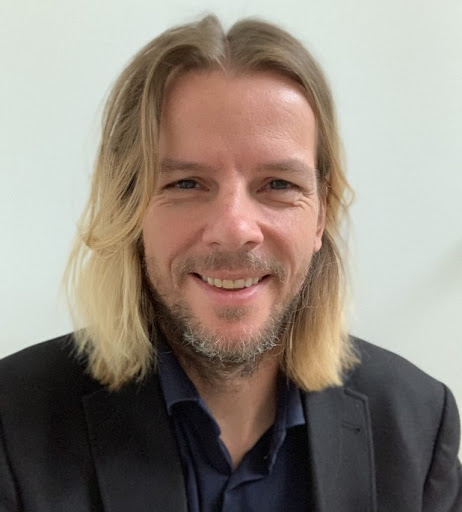}}]{Björn Schuller}
received his diploma, doctoral degree, habilitation, and Adjunct Teaching Professor in Machine Intelligence and Signal Processing all in EE/IT from TUM in Munich/Germany. He is Full Professor of Artificial Intelligence and the Head of GLAM at Imperial College London/UK, Full Professor and Chair of Embedded Intelligence for Health Care and Wellbeing at the University of Augsburg/Germany, and permanent Visiting Professor at HIT/China amongst other Professorships and Affiliations. Previous stays include Full Professor at the University of Passau/Germany, Researcher at Joanneum Research in Graz/Austria, and the CNRS-LIMSI in Orsay/France. He is a Fellow of the IEEE and Golden Core Awardee of the IEEE Computer Society, Fellow of the BCS, Fellow of the ISCA, President-Emeritus of the AAAC, and Senior Member of the ACM. He (co-)authored 1\,000+ publications (35k+ citations, h-index=85), is Field Chief Editor of Frontiers in Digital Health and was Editor in Chief of the IEEE Transactions on affective computing amongst manifold further commitments and service to the community. His 30+ awards include having been honoured as one of 40 extraordinary scientists under the age of 40 by the WEF in 2015. 
\end{IEEEbiography}

\end{document}